\documentclass[journal]{IEEEtran}

\usepackage[numbers,sort&compress]{natbib}
\usepackage{url}
\usepackage{bm}
\usepackage{amsfonts}
\usepackage{amsmath}
\usepackage{amssymb}
\usepackage{amsthm}
\usepackage{color}
\usepackage{enumerate}

\usepackage{array,color}
\usepackage{tabularx}
\usepackage{multirow}
\usepackage{booktabs}
\usepackage{makecell}  

\usepackage{graphicx}  
\usepackage{subfig}
\usepackage{epstopdf}
\usepackage{graphics}
\usepackage{epsfig}
\usepackage{psfrag}
\usepackage{overpic}

\usepackage{multicol}

 \usepackage{stfloats}
 \usepackage{float}   

 \usepackage{indentfirst} 
\usepackage{gensymb}

\usepackage{color, colortbl}

\begin{document}

\newcommand {\Data} [1]{\mbox{${#1}$}}  

\newcommand {\DataN} [2]{\Data{\Power{{#1}}{{{#2}}}}}  
\newcommand {\DataIJ} [3]{\Data{\Power{#1}{{{#2}\!\times{}\!{#3}}}}}  

\newcommand {\DatassI} [2]{\!\Data{\Index{#1}{\!\Data 1},\!\Index{#1}{\!\Data 2},\!\cdots,\!\Index{#1}{\!{#2}}}}  
\newcommand {\DatasI} [2]{\Data{\Index{#1}{\Data 1},\Index{#1}{\Data 2},\cdots,\Index{#1}{#2},\cdots}}   
\newcommand {\DatasII} [3]{\Data{\Index{#1}{{\Index{#2}{\Data 1}}},\Index{#1}{{\Index{#2}{\Data 2}}},\cdots,\Index{#1}{{\Index{#2}{#3}}},\cdots}}  

\newcommand {\DatasNTt}[3]{\Data{\Index{#1}{{#2}{\Data 1}},\Index{#1}{{#2}{\Data 2}},\cdots,\Index{#1}{{#2}{#3}}} } 
\newcommand {\DatasNTn}[3]{\Data{\Index{#1}{{\Data 1}{#3}},\Index{#1}{{\Data 2}{#3}},\cdots,\Index{#1}{{#2}{#3}}} } 

\newcommand {\Vector} [1]{\Data {\mathbf {#1}}}
\newcommand {\Rdata} [1]{\Data {\hat {#1}}}
\newcommand {\Tdata} [1]{\Data {\tilde {#1}}} 
\newcommand {\Udata} [1]{\Data {\overline {#1}}} 
\newcommand {\Fdata} [1]{\Data {\mathbb {#1}}} 
\newcommand {\Prod} [2]{\Data {\prod_{\SI {#1}}^{\SI {#2}}}}  
\newcommand {\Sum} [2]{\Data {\sum_{\SI {#1}}^{\SI {#2}}}}   
\newcommand {\Belong} [2]{\Data{ {#1} \in{}{#2}}}  

\newcommand {\Abs} [1]{\Data{ \lvert {#1} \rvert}}  
\newcommand {\Mul} [2]{\Data{ {#1} \times {#2}}}  
\newcommand {\Muls} [2]{\Data{ {#1} \! \times \!{#2}}}  
\newcommand {\Mulsd} [2]{\Data{ {#1} \! \cdot \!{#2}}}  
\newcommand {\Div} [2]{\Data{ \frac{#1}{#2}}}  
\newcommand {\Trend} [2]{\Data{ {#1}\rightarrow{#2}}}  
\newcommand {\Sqrt} [1]{\Data {\sqrt {#1}}} 
\newcommand {\Sqrnt} [2]{\Data {\sqrt[2]{#1}}} 

\newcommand {\Power} [2]{\Data{ {#1}^{\TI {#2}}}}  
\newcommand {\Index} [2]{\Data{ {#1}_{\TI {#2}}}}  

\newcommand {\Equ} [2]{\Data{ {#1} = {#2}}}  
\newcommand {\Equs} [2]{\Data{ {#1}\! =\! {#2}}}  
\newcommand {\Equss} [3]{\Equs {#1}{\Equs {#2}{#3}}}  

\newcommand {\Equu} [2]{\Data{ {#1} \equiv {#2}}}  

\newcommand {\LE}[0] {\leqslant}
\newcommand {\GE}[0] {\geqslant}
\newcommand {\NE}[0] {\neq}
\newcommand {\INF}[0] {\infty}
\newcommand {\MIN}[0] {\min}
\newcommand {\MAX}[0] {\max}

\newcommand {\SI}[1] {\small{#1}}
\newcommand {\TI}[1] {\tiny {#1}}
\newcommand {\Text}[1] {\text {#1}}

\newcommand {\VtS}[0]{\Index {t}{\Text {s}}}
\newcommand {\Vti}[0]{\Index {t}{i}}
\newcommand {\Vt}[0]{\Data {t}}
\newcommand {\VLES}[1]{\Index {\tau} {\SI{\Index {}{ \Text{#1}}}}}
\newcommand {\VLESmin}[1]{\Index {\tau} {\SI{\Index {}{ \Text{min\_\Text{#1}}}}}}
\newcommand {\VAT}[0]{\Index {\Vector A}{\Text{Time}}}
\newcommand {\VPbus}[1]{\Index {P}{\Text{Node-}{#1}}}
\newcommand {\VPbusmax}[1]{\Index {P}{\Text{max\_Node-}{#1}}}

\newcommand {\EtS}[2]{\Equs {\Index {t}{\Text {s}}}{#1} {#2}}
\newcommand {\Eti}[2]{\Equs {\Index {t}{i}}{#1} {#2}}
\newcommand {\Et}[2]{\Equs {t}{#1} {#2}}
\newcommand {\EMSR}[2]{\Equs {\Index {\tau} {\SI{\Index {}{ \Text{MSR}}}}}{#1} {#2}}
\newcommand {\EMSRmin}[2]{\Equs {\Index {\tau} {\SI{\Index {}{ \Text{MSR}}}}}{#1} {#2}}

\newcommand {\EAT}[2]{\Equs {\Index {\Vector A}{\Text{Time}}}{#1} {#2}}
\newcommand {\EPbus}[3]{\Equs {\Index {P}{\Text{Node-}{#1}}}{#2} \Text{ #3}}
\newcommand {\EPbusmax}[3]{\Equs {\Index {P}{\Text{max\_Node-}{#1}}}{#2} \Text{ #3}}

\newcommand {\Vgam}[1]{\Index {\gamma}{#1}}
\newcommand {\Egam}[2]{\Equs {\Vgam{#1}}{#2}}

\newcommand {\Emu}[2]{\Equs {{\mu}{#1}}{#2}}
\newcommand {\Esigg}[2]{\Equs {{\sigma}^2{#1}}{#2}}

\newcommand {\Vlambda}[1]{\Index {\lambda}{#1}}

\newcommand {\VV}[1]{\Index {\Vector V}{#1}}
\newcommand {\Vv}[1]{\Index {\Vector v}{#1}}
\newcommand {\Vsv}[1]{\Index {v}{#1}}
\newcommand {\VX}[1]{\Index {\Vector X}{#1}}
\newcommand {\VsX}[1]{\Index {X}{\SI{\Index {}{#1}}}}
\newcommand {\Vx}[1]{\Index {\Vector x}{\SI{\Index {}{#1}}}}
\newcommand {\Vsx}[1]{\Index {x}{\SI{\Index {}{#1}}}}
\newcommand {\VZ}[1]{\Index {\Vector Z}{#1}}
\newcommand {\Vz}[1]{\Index {\Vector z}{\SI{\Index {}{#1}}}}
\newcommand {\Vsz}[1]{\Index {z}{\SI{\Index {}{#1}}}}
\newcommand {\VIndex}[2]{\Index {\Vector {#1}}{#2}}
\newcommand {\VY}[1]{\Index {\Vector Y}{#1}}
\newcommand {\Vy}[1]{\Index {\Vector y}{#1}}
\newcommand {\Vsy}[1]{\Index {y}{\SI{\Index {}{#1}}}}

\newcommand {\VRV}[1]{\Index {\Rdata {\Vector V}}{#1}}
\newcommand {\VRsV}[1]{\Index {\Rdata {V}}{#1}}
\newcommand {\VRX}[1]{\Index {\Rdata {\Vector X}}{#1}}
\newcommand {\VRx}[1]{\Index {\Rdata {\Vector x}}{\SI{\Index {}{#1}}}}
\newcommand {\VRsx}[1]{\Index {\Rdata {x}}{\SI{\Index {}{#1}}}}
\newcommand {\VRZ}[1]{\Index {\Rdata {\Vector Z}}{#1}}
\newcommand {\VRz}[1]{\Index {\Rdata {\Vector z}}{\SI{\Index {}{#1}}}}
\newcommand {\VRsz}[1]{\Index {\Rdata {z}}{\SI{\Index {}{#1}}}}
\newcommand {\VTX}[1]{\Index {\Tdata {\Vector X}}{#1}}
\newcommand {\VTx}[1]{\Index {\Tdata {\Vector x}}{\SI{\Index {}{#1}}}}
\newcommand {\VTsx}[1]{\Index {\Tdata {x}}{\SI{\Index {}{#1}}}}
\newcommand {\VTsX}[1]{\Index {\Tdata {X}}{\SI{\Index {}{#1}}}}
\newcommand {\VTZ}[1]{\Index {\Tdata {\Vector Z}}{#1}}
\newcommand {\VTz}[1]{\Index {\Tdata {\Vector z}}{\SI{\Index {}{#1}}}}
\newcommand {\VTsz}[1]{\Index {\Tdata {z}}{\SI{\Index {}{#1}}}}
\newcommand {\VOG}[1]{\Vector{\Omega}{#1}}

\newcommand {\Sigg}[1]{\Data {{\sigma}^2({#1})}}
\newcommand {\Sig}[1]{\Data {{\sigma}({#1})}}

\newcommand {\Mu}[1]{\Data{{\mu} ({#1})}}
\newcommand {\Eig}[1]{\Data {\lambda}({\Vector {#1}}) }
\newcommand {\Her}[1]{\Power {#1}{\!H}}
\newcommand {\Tra}[1]{\Power {#1}{\!T}}

\newcommand {\VF}[3] {\DataIJ {\Fdata {#1}}{#2}{#3}}
\newcommand {\VRr}[2] {\DataN {\Fdata {#1}}{#2}}

\newcommand {\Tcol}[2] {\multicolumn{1}{#1}{#2} }
\newcommand {\Tcols}[3] {\multicolumn{#1}{#2}{#3} }
\newcommand {\Cur}[2] {\mbox {\Data {#1}-\Data {#2}}}

\newcommand {\VDelta}[1] {\Data {\Delta\!{#1}}}

\newcommand {\STE}[1] {\Fdata {E}{\Data{({#1})}}}
\newcommand {\STD}[1] {\Fdata {D}{\Data{({#1})}}}

\newcommand {\TestF}[1] {\Data {\varphi(#1)}}
\newcommand {\ROMAN}[1] {\uppercase\expandafter{\romannumeral#1}}

\title{Data-driven Estimation of the Power Flow Jacobian Matrix in High Dimensional Space}

\author{Xing~He, Lei~Chu, Robert~C. Qiu,~\IEEEmembership{Fellow,~IEEE},  Qian~Ai, Wentao~Huang

}

\maketitle


\begin{abstract}
The Jacobian matrix is the core part of power flow analysis, which is the basis for power system planning and operations. This paper estimates the Jacobian matrix in high dimensional space. Firstly, theoretical analysis and model-based calculation of the Jacobian matrix are introduced to obtain the benchmark value. Then, the estimation algorithms based on least-squared errors and the deviation estimation based on the neural network are studied in detail, including the theories, equations, derivations, codes, advantages and disadvantages, and application scenes.  The proposed algorithms are data-driven and sensitive to up-to-date topology parameters and state variables. The efforts are validate by comparing the results to benchmark values.
\end{abstract}


\IEEEpeerreviewmaketitle
\section{Introduction}
\label{Introd}
\IEEEPARstart {J}{acobian} matrix is a sparse matrix that results from a sensitivity analysis of
power flow equations. It is the key part of power flow analysis, which is the basis for power system planning and operations.
In additional, the eigenvalues of $\mathbf {J}$ have long been used as indices of system vulnerability to voltage instabilities \cite{GAO1992Voltage}. The sparsity structure of $\mathbf {J}$ inherently contains the most up-to-date network topology and corresponding parameters. Topology errors have long been cited as a cause of inaccurate state estimation results \cite{Lugtu1980Power, Wu1989Detection}.
Moreover, the Jacobian matrix, in practice, may be out-of-date due to erroneous records, faulty telemetry from remotely monitored circuit breakers, or unexpected operating conditions resulting from unforeseen equipment failure.

Our paradigms aim at supporting rapid Jacobian matrix estimation in a data rich but information limited environment, in the context of big data era \cite{he2015arch, Baek2015A}. These paradigms should effectively convert field data into information by allowing the power system operator to have a full understanding of the grid network operation. The proposed algorithm is data-driven and sensitive to up-to-date topology parameters and state variables.
For the most part of this article, we assume an offline model entirely unavailable and that all buses within the monitored region are equipped with data sensors such as PMUs, same as \cite{7364281}. We take the  node-to-ground admittance and normalization into account. The derivation is mathematically rigorous and matrix-based.

The proposed data-driven Jacobian estimation methods are validate by comparing the results to benchmark values obtained via direct linearization of the power flow equations at a particular operating point. The estimated Jacobian matrix is quite accurate and can therefore be used in studies that rely on the power flow model.

\section{Problem Formulation}
\label{Sec:ProFor}

\subsection{Grid Network Operation}
\begin{figure}[htpb]
\centering
\includegraphics[width=0.40\textwidth]{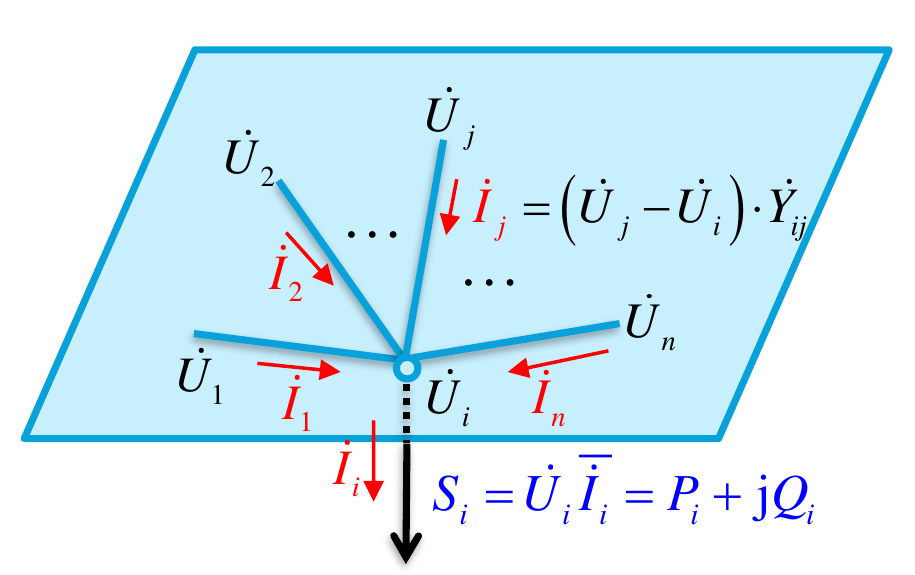}
\caption{Schematic Diagram  for Grid Network Operation}
\label{fig:GridOpr}
\end{figure}

For each node $i$ in a grid network, choosing the reference direction as shown in Fig \ref{fig:GridOpr}   , Kirchhoff's current law and ohm's law say that:
\begin{equation}
\label{eqI}
{{\dot{I}}_{i}}=\sum\limits_{\begin{smallmatrix}
 j=1 \\
 j\ne i
\end{smallmatrix}}^{n}{{{{\dot{I}}}_{j}}}=\sum\limits_{j\ne i}{{{{\dot{Y}}}_{ij}}\cdot \left( {{{\dot{U}}}_{j}}-{{{\dot{U}}}_{i}} \right)}.
\end{equation}
where ${\dot{Y}}_{ij}\!=\!G_{ij}\!+\!\mathrm{j} \cdot B_{ij}$  is the admittance in Cartesian form\footnote{$G$ is the conductance, $B$ is the susceptance, and $\mathrm{j}$ is the imaginary unit.}, and ${{\dot{U}}_{i}}\!=\!\left| {{{\dot{U}}}_{i}} \right|\angle {{\theta }_{i}}\!=\!{{V}_{i}}\angle {{\theta }_{i}}\!=\!{{V}_{i}}{{\text{e}}^{\text{j}{{\theta }_{i}}}}$ and ${{\dot{I}}_{i}}=\left| {{{\dot{I}}}_{i}} \right|\angle {{\phi }_{i}}$ are node voltage and node current, respectively, in polar form\footnote{${{V}_{i}}\angle {{\theta }_{i}}={{V}_{i}}{{\text{e}}^{\text{j}{{\theta }_{i}}}}={{V}_{i}}(\cos {{\theta }_{i}}+\text{j}\cdot \sin {{\theta }_{i}}).$}.


For all the nodes of the network, we obtain
\[
\left[ \begin{matrix}
   {{{\dot{I}}}_{1}}  \\
   \vdots   \\
   {{{\dot{I}}}_{n}}  \\
\end{matrix} \right]=\left[ \begin{matrix}
   {{y}_{11}} & {{y}_{12}} & \cdots  & {{y}_{1n}}  \\
   \vdots  & \vdots  & \cdots  & \vdots   \\
   {{y}_{n1}} & {{y}_{2n}} & \cdots  & {{y}_{nn}}  \\
\end{matrix} \right]\left[ \begin{matrix}
   {{{\dot{U}}}_{1}}  \\
   \vdots   \\
   {{{\dot{U}}}_{n}}  \\
\end{matrix} \right]
\]
or equivalently:
\begin{equation}
\label{eqII}
\mathbf{\dot{I}}=\mathbf{\dot{y}\dot{U}}
\end{equation}
where ${{[y]}_{ij}}=\left\{ \begin{aligned}
  & {{{\dot{Y}}}_{ij}}          &    i\ne j \\
 & -\sum\limits_{k\ne i}{{{{\dot{Y}}}_{ik}}}   &   i=j \\
\end{aligned} \right.$

\begin{equation}
\label{eqSS}
\begin{aligned}
  & \mathbf{\dot{S}}=\left[ \begin{matrix}
   {{P}_{1}}+\text{j}{{Q}_{1}}  \\
   {{P}_{2}}+\text{j}{{Q}_{2}}  \\
   \vdots   \\
   {{P}_{n}}+\text{j}{{Q}_{n}}  \\
\end{matrix} \right]=\dot{U}\circ \overline{{\dot{I}}}=\mathbf{\dot{U}}\circ \overline{\mathbf{Y\dot{U}}}=\mathbf{\dot{U}}\circ \overline{\mathbf{Y}}\overline{{\mathbf{\dot{U}}}} \\
 & =\sum{{{{\dot{U}}}_{i1}}{\mathbf{E}_{i1}}}\circ \sum{\overline{{{y}_{jk}}}{\mathbf{E}_{jk}}}\sum{\overline{{{{\dot{U}}}_{l1}}}{\mathbf{E}_{l1}}} \\
 & =\sum{{{{\dot{U}}}_{i1}}\overline{{{y}_{jk}}}\overline{{{{\dot{U}}}_{l1}}}{\mathbf{E}_{i1}}\circ \left( {\mathbf{E}_{jk}}{\mathbf{E}_{l1}} \right)}=\sum{{{{\dot{U}}}_{i}}\overline{{{y}_{ik}}}\overline{{{{\dot{U}}}_{k}}}{\mathbf{E}_{i1}}}
\end{aligned}
\end{equation}
where $\circ$ is the Hadamard product\footnote{$\mathbf C\!=\!\mathbf A\circ \mathbf B, [C]_{ij}\!=\![A]_{ij}[B]_{ij}.$}, and $\mathbf E_{ij}$ is the single-entry matrix---1 at $(i,j)$ and 0 elsewhere. It is worth mentioning that $\mathbf{A}\!=\!\sum\limits_{i,j}{{{\left[ A \right]}_{ij}}{\mathbf{E}_{ij}}}$, abbreviated as $\sum{{{A}_{ij}}{\mathbf{E}_{ij}}}$.

And thus, for each node $i$, its active power $P$ and reactive power $Q$ are expressed as:
\begin{equation}
\label{eq:PQend}
\begin{aligned}
  & \left\{ \begin{aligned}
  & {{P}_{i}}\!=\!{{V}_{i}}\sum\limits_{k\ne i}{{{V}_{k}}\left( {{G}_{ik}}\text{cos}{{\theta }_{ik}}\!+\!{{B}_{ik}}\text{sin}{{\theta }_{ik}} \right)}\!-\!{{V}_{i}}^{2}\sum\limits_{k\ne i}{{{G}_{ik}}} \\
 & {{Q}_{i}}\!=\!{{V}_{i}}\sum\limits_{k\ne i}{{{V}_{k}}\left( {{G}_{ik}}\sin {{\theta }_{ik}}\!-\!{{B}_{ik}}\text{cos}{{\theta }_{ik}} \right)}\!+\!{{V}_{i}}^{2}\sum\limits_{k\ne i}{{{B}_{ik}}} \\
\end{aligned} \right. \\
 &      \\
\end{aligned}
\end{equation}
For full details, see Eq. \eqref{eq:PQProc} in Appendix \ref{Sec:PQProc}.
Taking account of the $i$-th node-to-ground admittance  $y_i$,  we obtain
\begin{equation}
\label{eq:PQend1}
\left\{ \begin{aligned}
  & {{P}_{i}}:={{P}_{i}}-{{V}_{i}}^{2}{{g}_{i}} \\
 & {{Q}_{i}}:={{P}_{i}}+{{V}_{i}}^{2}{{b}_{i}} \\
\end{aligned} \right.
\end{equation}

Then we move to the core part of our study---Jacobian Matrix $\mathbf J.$
Abstractly, the physical power system obeying Eq. \eqref{eq:PQend1} may be viewed as an analog engine, taking bus voltage magnitudes $V$ and bus voltage phasers $\theta$ as \emph{inputs}, and "computing"  active power injection $P$ and reactive power injection $Q$  as \emph{outputs}.  Thus, the entries of $\mathbf J.$, i.e. ${\left[ J \right]}_{ij}$, are composed of partial derivatives of $P$ and $Q$ with respect to $V$ and $\theta$.  In all, $\mathbf J$ consists of four parts  $\mathbf H, \mathbf N, \mathbf K, \mathbf L$ as follows:
\begin{equation}
\label{eq:HNKLend}
\left\{ \begin{aligned}
  & {{H}_{ij}}\!={{V}_{i}}{{V}_{j}}\left( {{G}_{ij}}\sin {{\theta }_{ij}}\!-\!{{B}_{ij}}\cos {{\theta }_{ij}} \right)\!-\!{{\delta }_{ij}}\!\cdot\! {{Q}_{i}}\!+\!{{\delta }_{ij}}\!\cdot\! {V}_{i}^2 b_i \\
 & {{N}_{ij}}\!={{V}_{i}}{{V}_{j}}\left( {{G}_{ij}}\cos {{\theta }_{ij}}\!+\!{{B}_{ij}}\sin {{\theta }_{ij}} \right)\!+\!{{\delta }_{ij}}\!\cdot\! {{P}_{i}}\!-\!{{\delta }_{ij}}\!\cdot\! {V}_{i}^2 g_i \\
 & {{K}_{ij}}\!=-{{V}_{i}}{{V}_{j}}\left( {{G}_{ij}}\cos {{\theta }_{ij}}\!+\!{{B}_{ij}}\sin {{\theta }_{ij}} \right)\!+\!{{\delta }_{ij}}\!\cdot\! {{P}_{i}}\!+\!{{\delta }_{ij}}\!\cdot\! {V}_{i}^2 g_i \\
 & {{L}_{ij}}\!={{V}_{i}}{{V}_{j}}\left( {{G}_{ij}}\sin {{\theta }_{ij}}\!-\!{{B}_{ij}}\cos {{\theta }_{ij}} \right)\!+\!{{\delta }_{ij}}\!\cdot\! {{Q}_{i}}\!+\!{{\delta }_{ij}}\!\cdot\! {V}_{i}^2 b_i \\
\end{aligned} \right.
\end{equation}
where $ {{H}_{ij}}\!=\!\frac{\partial {{P}_{i}}}{\partial {{\theta }_{j}}}, {{N}_{ij}}\!=\!\frac{\partial {{P}_{i}}}{\partial {{V}_{j}}}{{V}_{j}}, {{K}_{ij}}\!=\!\frac{\partial {{Q}_{i}}}{\partial {{\theta }_{j}}}, {{L}_{ij}}\!=\!\frac{\partial {{Q}_{i}}}{\partial {{V}_{j}}}{{V}_{j}}$
, and ${\delta}$ is the Kronecker Delta Function defined as
\[
\delta_{\alpha,\beta}=
\begin{cases}
    1 & \alpha=\beta \\
    0 & \alpha\neq\beta.
\end{cases}
\]
For full details, see Appendix \ref{Sec:HNKLProc}.

\subsection{Power Flow Analysis  and Theoretical Calculation}
Power Flow (PF) analysis is the fundamental and most heavily used tool for solving many complex power system operation problems, such as fault diagnosis, state estimation, $N\!-\!1$ security, optimal power dispatch, etc.
PF analysis deals mainly with the calculation of the steady-state voltage magnitude  and phase  for each network bus, for a given set of variables such as load demand and real power generation, under certain assumptions such as balanced system operation \cite{gomez2018electric}. Based on this information, the network operating conditions, in particular, real and reactive power flows on each branch, power loesses, and generator reactive power outputs, can be determined \cite{Vaccaro2017A}. Thus, the input (output) variables of the PF problem typically fall into three categories:
\begin{itemize}
\item active power $P$ and voltage magnitude $V$ (reactive power $Q$ and voltage angle $\theta$) for each voltage controlled bus, i.e. $PV$ buses;
\item active power $P$ and reactive power $Q$ (voltage magnitude $V$ and voltage angle $\theta$) for each load bus, i.e. $PQ$ buses;
\item voltage magnitude $V$ and voltage angle $\theta$ (active power $P$ and reactive power $Q$) for reference or slack bus.
\end{itemize}

Theoretical calculation for PF analysis is model-based and assumption-based. That is to say, prior information of topological parameter, the admittance $Y$,  is required in advance.
Consider a power system with $n$ buses among which there are  $m$ $PV$ buses,  $n\!-\!m\!-\!1$  $PQ$ buses, and $1$ slack bus. Under the background of Eq. \eqref{eq:HNKLend}, the PF problem can be formalized in general as Eq. \eqref{eq:J1}, which simultaneously solves a set of equations with an equal number of unknowns \cite{JD2012PowerDesign}.

For  Eq. \eqref{eq:J1}, $\bm f$ is a differentiable mapping, $\bm f\!:\!\mathbf{x}\!\in\! {{\mathbb{R}}^{2n-m-2}}\!\to\! \mathbf{y}\!\in\! {{\mathbb{R}}^{2n-m-2}}$, and $\mathbf{J}$ is the Jacobian matrix,
\begin{equation}
\label{eq:J1}
\mathbf{y}\!\triangleq\! \left[ \begin{matrix}
   {{P}_{1}}  \\
   \vdots   \\
   {{P}_{n-1}}  \\
   {{Q}_{m+1}}  \\
   \vdots   \\
   {{Q}_{n-1}}  \\
\end{matrix} \right]\!=\!\bm f\left[ \begin{matrix}
   {{\theta }_{1}}  \\
   \vdots   \\
   {{\theta }_{n-1}}  \\
   {{V}_{m+1}}  \\
   \vdots   \\
   {{V}_{n-1}}  \\
\end{matrix} \right]\!\triangleq\!\bm f\left( \mathbf{x} \right)
\qquad \mathbf{J}\!=\!\left[ \begin{matrix}
   \frac{\partial {{y}_{1}}}{\partial {{x}_{1}}} & \cdots  & \frac{\partial {{y}_{1}}}{\partial {{x}_{l}}}  \\
   \vdots  & \ddots  & \vdots   \\
   \frac{\partial {{y}_{l}}}{\partial {{x}_{1}}} & \cdots  & \frac{\partial {{y}_{l}}}{\partial {{x}_{l}}}  \\
\end{matrix} \right]
\end{equation}
According to Eq. \eqref{eq:HNKLend}, $\mathbf J$ is calculated
\begin{equation}
\label{eq:JHNKL}
\mathbf{J}= \left[ \begin{array}{*{35}{l}}
   {{\left[ \mathbf H \right]}_{n-1, n-1}} & {{\left[ \mathbf N \right]}_{n-1, n-m-1}}  \\
   {{\left[ \mathbf K \right]}_{n-m-1, n-1}} & {{\left[ \mathbf L \right]}_{n-m-1, n-m-1}}  \\
\end{array} \right]
\end{equation}

Numerical iteration algorithms and sparse factorization techniques, mainly based on Newton-Raphson or fast-decoupled methods, are employed to approximate the nonlinear
PF equations by linearized  $\mathbf J$.
In mathematics, if $\mathbf p$ is a point in $\mathbb{R}^n$ and $\bm f$ is differentiable at  $\mathbf p$ , then its derivative is given by $\mathbf {J}_{\bm {f} }(\mathbf {p}) $. In this case, the linear map described by $\mathbf {J} _{\bm {f} }(\mathbf {p} )$ is the best linear approximation of $\bm f$  near the point $\mathbf p$, in the sense that
\begin{equation}
\label{eq:TylorV1}
{\bm {f} (\mathbf {b} )= \bm {f} (\mathbf {p} )+\mathbf {J} _{\bm {f} }(\mathbf {p} )(\mathbf {b} -\mathbf {p} )+o(\|\mathbf {b} -\mathbf {p} \|)}
\end{equation}
for $\mathbf {b}$ close to $\mathbf {p}$ and where $o$ is the little $o$-notation (for $\mathbf {b} \to \mathbf {p}$) and $\left\| \mathbf{b}-\mathbf{p} \right\|$ is the distance between $\mathbf {b}$ and $\mathbf {p}$.


From Eq. \eqref{eq:TylorV1}, the linear approximation of the system operating from point $(\mathbf{x}^{(k)},\mathbf{y}^{(k)})$ to point $(\mathbf{x}^{(k\!+\!1)},\mathbf{y}^{(k\!+\!1)})$, the iteration is acquired as follows:
\begin{equation}
\label{eq:XJ}
{{\mathbf{x}}^{\left( k+1 \right)}}:={{\mathbf{x}}^{\left( k \right)}}+{\mathbf {J}_{\bm{f}}}^{-1}\left( {{\mathbf{x}}^{\left( k \right)}} \right)\left( {{\mathbf{y}}^{\left( k+1 \right)}}-{{\mathbf{y}}^{\left( k \right)}} \right)
\end{equation}
where $:=$ is the assignment symbol in computer science.

The iteration, given in Eq.  \eqref{eq:XJ}, depicts how the power system state estimation is carried out.
 ${\mathbf{y}}^{\left( k+1 \right)}$, according to Eq. \eqref{eq:J1}, is the desired $P, Q$ of the $PQ$ buses and desired $P$ of the $PV$ buses\footnote{For $PQ$ buses, neither $V$ and $\theta$ are fixed; they are state variables that need to be estimated.  For $PV$ buses, $V$ is fixed, and $\theta$ are state variables that need to be estimated.}.  ${\mathbf{y}}^{\left( k \right)}$ and ${\mathbf{x}}^{\left( k \right)}$ are measurements available from sensors.  In order to conduct Eq.  \eqref{eq:XJ},  Jacobian Matrix $\mathbf {J}$ is acquired.

Traditionally, $\mathbf {J}$ is computed offline via Eq. \eqref{eq:JHNKL} based on the physical model of the network.
The model-based approach is not ideal in practice, since accurate and up-to-date network topology and relevant parameters (the admittance $Y$)
and operating point $(\mathbf{x}^{(k)},\mathbf{y}^{(k)})$ are required according to Eq. \eqref{eq:HNKLend}.

\section{Data-driven Approach to Jacobian Matrix Estimation}
This section
proposes a data-driven  estimation of the Jacobian Matrix $\mathbf {J}$.
In the data-driven mode, the aforementioned  physical model and admittance $Y$ are no longer necessary information.
The data-driven approach could handle  the scenario in which the system topology and parameter information are wholly unavailable.
Moreover, the estimation of the Jacobian Matrix $\mathbf {J}$ inherently contains the most up-to-date network topology and corresponding parameters.

\subsection{Data-driven View of Grid Operation and its Matrix Form}
Under fairly general conditions,  our estimation target $\mathbf {J}$  is almost unchange within a short time, e.g. $\Delta t$, due to the stability of the system, or concretely, variable $V, \theta, Y$ according to Eq. \eqref{eq:HNKLend}. During  $\Delta t$, considering $T$ times observation at time instants $t_i$, $(i\!=\!1,2,\cdots,T, t_T\!-\!t_1\!=\!\Delta t)$, operating points $(\mathbf{x}^{(i)},\mathbf{y}^{(i)})$ are obtained as  Eq. \eqref{eq:J1}. Defining $\Delta {{\mathbf{x}}^{\left( k \right)}}\!\triangleq\! {{\mathbf{x}}^{\left( k+1 \right)}}\!-\!{{\mathbf{x}}^{\left( k \right)}}$, and similarly $\Delta {{\mathbf{y}}^{\left( k \right)}}\!\triangleq\! {{\mathbf{y}}^{\left( k+1 \right)}}\!-\!{{\mathbf{y}}^{\left( k \right)}}$, from Eq.  \eqref{eq:TylorV1} we deduce that $\Delta {{\mathbf{y}}^{\left( k \right)}}\!\approx\!\mathbf {J}\Delta {{\mathbf{x}}^{\left( k \right)}}$. Since $\mathbf {J}$ keeps nearly constant, the matrix form is written as:
\begin{equation}
\label{Eq:MMYX}
\mathbf B \!\approx\! \mathbf J\mathbf{A}
\end{equation}
where $\mathbf{J} \!\in\! {{\mathbb{R}}^{N\!\times\!N}} $ ($N\!=\!2n-m-2$), $\mathbf{B} \!=\! \left[ {\Delta {{{\mathbf{y}}^{(1)}}} , \cdots ,\Delta  {{{\mathbf{y}}^{(T)}}}} \right] \!\in\! {{\mathbb{R}}^{N\!\times\!T}}$, and $\mathbf{A}\! =\! \left[ {\Delta {{{\mathbf{x}}^{(1)}}} , \cdots ,\Delta  {{{\mathbf{x}}^{(T)}}}} \right] \!\in\! {{\mathbb{R}}^{N\!\times\!T}}$. As the fast-sampling of PMUs, it is reasonable to assume that  $T\!>\!N\!$.

It is worth mentioning that sometimes the normalization is required. Supposing that $\mathbf{\overset{\scriptscriptstyle\frown}{B}}\!=\!{{\mathbf{\Lambda }}_{\mathbf{B}}}\mathbf{B}$, $\mathbf{\overset{\scriptscriptstyle\frown}{A}}\!=\!{{\mathbf{\Lambda }}_{\mathbf{A}}}\mathbf{A}$, and $\mathbf{\overset{\scriptscriptstyle\frown}{B}}\!\approx\! \mathbf{\overset{\scriptscriptstyle\frown}{J}\overset{\scriptscriptstyle\frown}{A}}$, it is deduced that  $\mathbf{J}\!=\!{{\mathbf{\Lambda }}_{\mathbf{B}}}^{-1}\mathbf{\overset{\scriptscriptstyle\frown}{J}}{{\mathbf{\Lambda }}_{\mathbf{A}}}.$

\subsection{Least Squares Method}
From   Eq. \eqref{Eq:MMYX} we deduce that ${{\mathbf{B}}^{\text{T}}}\!\approx\! {{\mathbf{A}}^{\text{T}}}{{\mathbf{J}}^{\text{T}}}$. Therefore,
\begin{equation}
\label{Eq:VecYX}
\bm{\beta}_i \approx \bm{\Lambda }\bm{\vartheta}_i
\end{equation}
where  $\bm{\beta}_i\!\in\! {{\mathbb{R}}^{T}}$ is the $i$-th column of $\mathbf{B}^{\text{T}}$,  $\bm{\vartheta}_i$ of $\mathbf{J}^{\text{T}}$, and matrix $ \bm{\Lambda }\!\!\triangleq\!{\mathbf{A}}^{\text{T}}\!\in\! {{\mathbb{R}}^{T\!\times\!N}}$ is an overdetermined matrix due to $T\!>\!N\!$.
\subsubsection{Ordinary LSE Estimation}
{\Text{\\}}

In ordinary least-squares errors (LSE) estimation,  the regressor matrix $\bm \Lambda$ is assumed to be error free.
The rationale is to correct the observations $\bm{\beta}_i$ as little as possible under the Euclidean norm metric \cite{van1991total}; this
can be formulated as an optimization problem:
\[
  \underset{{{{\hat{\bm \vartheta }}}_{i}}\in {{\mathbb{R}}^{N}}}{\mathop{\arg \min }}\, {{\left\| \bm \beta_{i} - \bm \Lambda \hat{\bm \vartheta_{i}} \right\|}_{2}} \\
\]
We assume $\bm \Lambda$ has full column rank (rank $N$); under this condition, the closed-form unique solution is
\[
\hat{\bm \vartheta_{i}} ={{\left( {{\bm \Lambda }^{\text{T}}}\bm \Lambda  \right)}^{-1}}{{\bm \Lambda }^{\text{T}}}\bm \beta_{i}
\]
The proof is given in Appendix \ref{Sec:MatrixOpr}.

Furthermore, the estimation for $\mathbf {J}$ is
\begin{equation}
\label{Eq:LSEsol}
 {\hat{\mathbf{J}}^{\text{T}}}={{\left( \mathbf{A}{{\mathbf{A}}^{\text{T}}} \right)}^{-1}}\mathbf{A}{{\mathbf{B}}^{\text{T}}}
\end{equation}
and the Octave/Matlab code is:\\
\begin{normalsize}
\begin{small}
{\color{blue} function} JF = lse(A, B)\\
JT = pinv(A*A')*A*B';  { } JF = JT'; \\
{\color{blue} end}
\end{small}
\end{normalsize}
\subsubsection{Total Least Squares}
{\Text{\\}}

In practice, measurement and modeling errors enter into both the regressor matrix and the observation vectors.
Taking the errors of  $\bm \Lambda$  into account, we face a Total Least Squares (TLS) problem:
\[
\bm{\Theta} \approx \bm{\Lambda }\mathbf{J}^{\text{T}},
\]
where  $\bm{\Theta}\!\triangleq\!\mathbf{B}^{\text{T}}\!\in\! {{\mathbb{R}}^{T\!\times\!N}}$.

According to \cite{wiki2018TLS}, the above problem is solved via singular value decomposition (SVD).
We seek to find a $\mathbf{J}$ that minimizes error matrices $\mathbf E$ and $\mathbf F$, respectively,  for $\bm{\Lambda}$  and $\bm{\Theta}$:
\begin{equation}
\label{Eq:TLSJ}
\underset{\mathbf{J}}{\mathop{\text{argmin}}}\,{{\left\| \left[ \begin{matrix}   \mathbf{E} & \mathbf{F}  \\ \end{matrix} \right] \right\|}_{\text F}}  \quad ,\quad \left(\bm{\Lambda}+\mathbf{E} \right)\mathbf{J}^{\text{T}}\!=\!\bm{\Theta}+\mathbf{F}
\end{equation}
where $\left[ \begin{matrix}
   \mathbf{E} & \mathbf{F}  \\
\end{matrix} \right]$ is the augmented matrix with $\mathbf{E}$ and $\mathbf{F}$ side by side and ${\displaystyle \|\cdot \|_{\text F}} $ is the Frobenius norm.

This can be rewritten as
\[{ [\begin{matrix} (\bm{\Lambda}+\mathbf E)&(\bm{\Theta}+\mathbf F)\end{matrix} ]{\begin{bmatrix}\mathbf{J}^{\text{T}}\\-\mathbf I_{k}\end{bmatrix}}=\mathbf 0.} \]
where $\mathbf I_{k}$ is the ${k\times k}$ ($k\!:=\!N$) identity matrix. The goal is then to find $\left[ \begin{matrix}   \mathbf{E} & \mathbf{F}  \\ \end{matrix} \right]$ that reduces the rank of $\left[ \begin{matrix}   {\bm{\Lambda}} & {\bm{\Theta}}  \\ \end{matrix} \right]$ by $k$.
Conduct SVD of both $\left[ \begin{matrix}   {\bm{\Lambda}} & {\bm{\Theta}}  \\ \end{matrix} \right]$ and $[\begin{matrix} (\bm{\Lambda}+\mathbf E)&(\bm{\Theta}+\mathbf F)\end{matrix} ]$:
\normalsize{}
\small{}
\begin{equation}
\label{Eq:svd12}
\left\{ \begin{aligned}
   \left[ \begin{matrix}   {\bm{\Lambda}} & {\bm{\Theta}}  \\ \end{matrix} \right] =& {{\mathbf{U}}^{\left( 1 \right)}}{{\bm{\Sigma }}^{\left( 1 \right)}}{{\mathbf{V}}^{\left( 1 \right)\text{T}}} &\qquad (1)
 \\
  [(\bm{\Lambda}+\mathbf E)\;(\bm{\Theta}+\mathbf F)] =&  {{\mathbf{U}}^{\left( 2 \right)}}{{\bm{\Sigma }}^{\left( 2 \right)}}{{\mathbf{V}}^{\left( 2 \right)\text{T}}} &\qquad (2)
\\
\end{aligned} \right.,
\end{equation}
where $\mathbf{U}, \mathbf{V}$ are unitary matrix and  $\bm \Sigma$ is  diagonal matrix. Furthermore
\[{{\mathbf{U}}^{\left( i \right)}}{{\mathbf{\Sigma }}^{\left( i \right)}}{{\mathbf{V}}^{\left( i \right)\text{T}}}\!\triangleq\! \left[ \begin{matrix}
   {{\mathbf{U}}_{1}}^{\left( i \right)} & {{\mathbf{U}}_{2}}^{\left( i \right)}  \\
\end{matrix} \right]\left[ \begin{matrix}
   {{\mathbf{\Sigma }}_{1}}^{\left( i \right)} & \mathbf{0}  \\
   \mathbf{0} & {{\mathbf{\Sigma }}_{2}}^{\left( i \right)}  \\
\end{matrix} \right]\left[ \begin{matrix}
   {{\mathbf{V}}_{11}}^{\left( i \right)} & {{\mathbf{V}}_{12}}^{\left( i \right)}  \\
   {{\mathbf{V}}_{21}}^{\left( i \right)} & {{\mathbf{V}}_{22}}^{\left( i \right)}  \\
\end{matrix} \right]^{\text{T}}
\]
\normalsize{}
where $i\!=\!1,2$ and $\mathbf{U}, \bm{\Sigma}, \mathbf{V}$ are partitioned into blocks corresponding to the shape of $\mathbf{\Lambda }$ and $\mathbf{\Theta}$.
It is worth mentioning that $\mathbf{\Lambda }\!=\!\mathbf{U}_{\mathbf{\Lambda }}\bm \Sigma_{\mathbf{\Lambda }}\mathbf{V}_{\mathbf{\Lambda }}^\text{T}$ and $\mathbf{\Theta }\!=\!\mathbf{U}_{\mathbf{\Theta }}\bm \Sigma_{\mathbf{\Theta }}\mathbf{V}_{\mathbf{\Theta }}^\text{T}$.

Using the Eckart--Young Theorem, the approximation minimising the norm of the error is such that matrices $\mathbf{U}$ and $\mathbf{V}$ are unchanged, while the  $k$-smallest singular values are replaced with 0. With the characteristics of the SVD for augmented matrix\footnote{Details  are given in Appendix \ref{Sec:SVDXYZ}.}, we obtained that   ${{\mathbf{U}}_{1}}^{\left( 1 \right)}\!=\!{{\mathbf{U}}_{1}}^{\left( 2 \right)}\!=\!{{\mathbf{U}}_{{\mathbf{\Lambda }}}}\!, {{\mathbf{U}}_{2}}^{\left( 1 \right)}\!=\!{{\mathbf{U}}_{2}}^{\left( 2 \right)}\!=\!{{\mathbf{U}}_{{\mathbf{\Theta }}}}\!,{\bm{\Sigma}_{1}}^{\left( 1 \right)}\!=\!{\bm{\Sigma}_{1}}^{\left( 2 \right)}\!=\!{\bm{\Sigma}_{{\mathbf{\Lambda }}}}\!,{\bm{\Sigma}_{2}}^{\left( 1 \right)}\!=\!{\bm{\Sigma}_{{\mathbf{\Theta }}}}\!, {{\mathbf{\Sigma }}_{2}}^{\left( 2 \right)}\!=\!\bm 0$, and ${\mathbf{V}}^{\left( 1 \right)}\!=\!{\mathbf{V}}^{\left( 2 \right)}\!=\!\left[ \begin{matrix}
   {{\mathbf{V}}_{11}} & {{\mathbf{V}}_{12}}  \\
   {{\mathbf{V}}_{21}} & {{\mathbf{V}}_{22}}  \\
\end{matrix} \right]
\!\triangleq\! \left[ \begin{matrix}
   {{\mathbf{V}}_{:1}} & {{\mathbf{V}}_{:2}}  \\
\end{matrix} \right]
$ are deterministic, although unknown.

Staring with Eq. \eqref{Eq:svd12}, $(1)-(2)$:
\begin{equation}
\label{eq:EFaug}
\left[ \begin{matrix}   \mathbf{E} & \mathbf{F}  \\ \end{matrix} \right]\!=\!-\left[ \begin{matrix}
   {{\mathbf{U}}_{\mathbf{\Lambda }}} & {{\mathbf{U}}_{\mathbf{\Theta }}}  \\
\end{matrix} \right]\left[ \begin{matrix}
   \mathbf{0} & \mathbf{0}  \\
   \mathbf{0} & {{\mathbf{\Sigma }}_{\mathbf{\Theta }}}  \\
\end{matrix} \right]
\left[ \begin{matrix}
   {{\mathbf{V}}_{:1}} & {{\mathbf{V}}_{:2}}  \\
\end{matrix} \right]
^{\text{T}}
\end{equation}

Note that
\[
   \left[ \begin{matrix}
   \mathbf{0} & \mathbf{0}  \\
   \mathbf{0} & {{\mathbf{\Sigma }}_{\mathbf{\Theta }}}  \\
\end{matrix} \right]
 =\left[ \begin{matrix}
   {{\mathbf{\Sigma }}_{\mathbf{\Lambda }}} & \mathbf{0}  \\
   \mathbf{0} & {{\mathbf{\Sigma }}_{\mathbf{\Theta }}}  \\
\end{matrix} \right]{{\left[ \begin{matrix}
   {{\mathbf{V}}_{:1}} & {{\mathbf{V}}_{:2}}  \\
\end{matrix} \right]}^{\text{T}}}\left[ \begin{matrix}
   \mathbf{0} & {{\mathbf{V}}_{:2}}  \\
\end{matrix} \right]
\]
As a result, Eq. \eqref{eq:EFaug} is turned into:
\begin{normalsize}
\begin{small}
\[
\begin{aligned}
  & \left[ \begin{matrix}
   \mathbf{E} & \mathbf{F}  \\
\end{matrix} \right]\!=\!-\left[ \begin{matrix}
   \mathbf{\Lambda } & \mathbf{\Theta }  \\
\end{matrix} \right]\!\left[ \begin{matrix}
   \mathbf{0} & {{\mathbf{V}}_{:2}}  \\
\end{matrix} \right]\!{{\left[ \begin{matrix}
   {{\mathbf{V}}_{:1}} & {{\mathbf{V}}_{:2}}  \\
\end{matrix} \right]}^{\text{T}}}\!=\!-\!\left[ \begin{matrix}
   \mathbf{\Lambda } & \mathbf{\Theta }  \\
\end{matrix} \right]{{\mathbf{V}}_{:2}}{{\mathbf{V}}_{:2}}^{\text{T}} \\
 &\Rightarrow \left[ \begin{matrix}
   \mathbf{E} & \mathbf{F}  \\
\end{matrix} \right]{{\mathbf{V}}_{:2}}\!=\!-\!\left[ \begin{matrix}
   \mathbf{\Lambda } & \mathbf{\Theta }  \\
\end{matrix} \right]{{\mathbf{V}}_{:2}}\mathbf{I}
\Rightarrow \left[ \begin{matrix}
   \mathbf{\Lambda }\!+\!\mathbf{E} & \mathbf{\Theta }\!+\!\mathbf{F}  \\
\end{matrix} \right]\left[ \begin{matrix}
   {{\mathbf{V}}_{12}}  \\
   {{\mathbf{V}}_{22}}  \\
\end{matrix} \right]=\mathbf{0} \\
\end{aligned}
\]
\end{small}
\end{normalsize}

If ${{\mathbf{V}}_{22}} $ is nonsingular\footnote{the behavior of TLS when ${{\mathbf{V}}_{22}} $ is singular is not well understood yet}, we can then right multiply both sides by $-{{\mathbf{V}}_{22}}^{{-1}}$ to bring the bottom block of the right matrix to the negative identity, giving:
\[\left[ \begin{matrix}
   \mathbf{\Lambda }+\mathbf{E} & \mathbf{\Theta }+\mathbf{F}  \\
\end{matrix} \right]\left[ \begin{matrix}
   -{{\mathbf{V}}_{12}}{{\mathbf{V}}_{22}}^{-1}  \\
   -\mathbf{I}  \\
\end{matrix} \right]=\mathbf{0}\]
and so
\[{{\mathbf{J}}^{\text{T}}}=-{{\mathbf{V}}_{12}}{{\mathbf{V}}_{22}}^{-1}\]
and the Octave/Matlab code is:\\
\begin{normalsize}
\begin{small}
{\color{blue} function} JF = tls(A, B)\\
X = A'; { }Y = B';\\
{[m n]} = size(X); { } Z = [X Y]; \\
{[U S V]} = svd(Z, 0); \\          
V12     = V(1:n, 1+n:end); { }V22  = V(1+n:end, 1+n:end); \\    
JT      = -V12*pinv(V22);  { }  JF = JT';\\
{\color{blue} end}
\end{small}
\end{normalsize}


\subsubsection{Weighted Least Squares}
{\Text{\\}}

For some scenarios the observations may be weighted---for example, they may not be equally reliable due to data quality. In this case, one can minimize the weighted sum of squares starting with Eq. \eqref{Eq:VecYX} $\bm{\beta}_i \approx \bm{\Lambda }\bm{\vartheta}_i$:
\[\underset{\vartheta }{\mathop{\arg \min }}\,\sum\limits_{k=1}^{T}{{{w}_{k}}}{{\left| {{\beta }_{ik}}-\mathbf{\Lambda }{{\vartheta }_{k}} \right|}^{2}}\!=\!\underset{\vartheta }{\mathop{\arg \min }}\,{{\left\| {{\mathbf{W}}^{1/2}}\left( {{\beta }_{i}}-\mathbf{\Lambda }\vartheta  \right) \right\|}^{2}}\]
It is a typical Weighted least squares (WLS) problem.
The closed-form unique solution is
\[{\hat{\bm \vartheta }_{i}}={{\left( {{\mathbf{\Lambda }}^{\text{T}}}\mathbf{W}\mathbf{\Lambda } \right)}^{-1}}{{\mathbf{\Lambda }}^{\text{T}}}\mathbf{W}\bm \beta_{i}\]

Furthermore, the estimation for $\mathbf {J}$ is
\[{\hat{\mathbf{J}}^{\text{T}}}={{\left( {{\mathbf{A}}}\mathbf{W}\mathbf{A}^{\text{T}} \right)}^{-1}}{{\mathbf{A}}}\mathbf{W}{{\mathbf{B}}^{\text{T}}}\]

\subsubsection{Generalized Least Squares}
{\Text{\\}}

The aforementioned WLS is a special case of generalized least squares (GLS) occurring when ${{\mathbf{\Omega }}^{-1}}:=\mathbf{W}$.
This situation arises when the variances of the observed values are unequal (i.e. heteroscedasticity is present), but where no correlations exist among the observed variances. The weight for unit $i$ is proportional to the reciprocal of the variance of the response for unit $i$ \cite{strutz2010data}; hence all the off-diagonal entries of $\mathbf \Omega$ are 0.


Starting with LSE model Eq. \eqref{Eq:VecYX} $\bm{\beta}_i \approx \bm{\Lambda }\bm{\vartheta}_i$, GLS forces the conditional mean of $\bm \beta$  given $\bm{\Lambda }$  to be a linear function of $\bm{\Lambda }$, and assumes the conditional variance of the error term given $\bm{\Lambda }$    is a known nonsingular covariance matrix $\mathbf{\Omega}$. This is usually written as
\begin{equation}
\label{eq:gls1}
{{\bm \beta }_{i}}=\mathbf{\Lambda }{{\bm \vartheta }_{i}}+{{\bm \varepsilon }_{i}},  \qquad \mathbb{E}\left[\bm  \varepsilon \left| \mathbf{\Lambda } \right. \right]=\mathbf{0}, \quad \operatorname{cov}\left[\bm  \varepsilon \left| \mathbf{\Lambda } \right. \right]=\mathbf{\Omega }
\end{equation}
Here ${\bm  \vartheta_i  \in \mathbb {R} ^{k}}$ is a vector of unknown constants (known as regression coefficients) that must be estimated from the data.
The generalized least squares method estimates $\bm  \vartheta_i $ by minimizing the squared Mahalanobis length of this residual vector:
\[\underset{\hat{{\bm \vartheta }_{i}}}{\mathop{\arg \min }}\,{{\left( {\bm {\beta }_{i}}-\mathbf{\Lambda }\hat{{\bm \vartheta }_{i}} \right)}^{\text{T}}}{{\mathbf{\Omega }}^{-1}}\left( {{\bm \beta }_{i}}-\mathbf{\Lambda }\hat{{\bm \vartheta }_{i}} \right)\]
Since the objective is a quadratic form in $\hat {\bm  \vartheta}_i $, the estimator has an explicit formula:
\begin{equation}
\label{eq:pd4}
{\hat{\bm \vartheta }_{i}}={{\left( {{\mathbf{\Lambda }}^{\text{T}}} {{\mathbf{\Omega }}^{-1}}\mathbf{\Lambda } \right)}^{-1}}{{\mathbf{\Lambda }}^{\text{T}}}{{\mathbf{\Omega }}^{-1}}\bm \beta_{i}
\end{equation}
and then,
\begin{equation}
\label{eq:pd5}
\hat{\mathbf J}^{\text{T}}=\left( \mathbf{A} {{\mathbf{\Omega }}^{-1}} \mathbf{A}^{\text{T}} \right)^{-1}\mathbf{A}\mathbf{\Omega }^{-1}\mathbf {B}^{\text{T}}
\end{equation}

GLS estimator could be conducted  to obtain the parameters $\mathbf{\Omega }$  through iteration. As aforementioned in Eq. \eqref{eq:gls1} and \eqref{eq:pd5}, we give the Octave/Matlab code for the iteration as follows:\\
\begin{normalsize}
\begin{small}
{\color{blue} function} JF = gls(A, B)\\
$[$N,T$]$ = size(A); { } G = eye(T);  { }  Iter = 100;\\
{\color{blue}for} ti = 1:Iter
\par \setlength{\parindent}{2em} invG = pinv(G); { } JT  = pinv(A*invG*A')*A*invG*B';
\par \setlength{\parindent}{2em} E = B - JT'*A; { } E = E';
\par \setlength{\parindent}{2em} G = 1/N*E*E';  { }   G = diag(diag(G));\\
{\color{blue}end}\\
JF = JT';\\
{\color{blue} end}
\end{small}
\end{normalsize}

\subsection{Estimation via Kronecker product}

The Kronecker product \cite{wiki2018kron} can be used to get a convenient representation for some matrix equations. Consider equation $\mathbf {AXB}\!=\!\mathbf {C}$, where $\mathbf A$, $\mathbf B$ and $\mathbf C$ are given matrices and matrix $\mathbf X$ is the unknown. We can rewrite this equation as
\[
{\displaystyle \left(\mathbf {B} ^{\textsf {T}}\otimes \mathbf {A} \right)\,\operatorname {vec} (\mathbf {X} )=\operatorname {vec} (\mathbf {AXB} )=\operatorname {vec} (\mathbf {C} ).}
\]
Here, $\operatorname {vec} (\mathbf {X} )$ denotes the vectorization of the matrix X formed by stacking the columns of X into a single column vector.


Rewriting Eq. \eqref{Eq:MMYX}: $\mathbf B \!\approx\! \mathbf J\mathbf{A}$, we obtain
\[\operatorname{vec}(\mathbf{B})\approx \operatorname{vec}({{\mathbf{I}}_{N}}\mathbf{JA})=\left( {{\mathbf{A}}^{\text{T}}}\otimes {{\mathbf{I}}_{N}} \right)\operatorname{vec}(\mathbf{J})\triangleq \mathbf{T}\operatorname{vec}(\mathbf{J})\]
If $\mathbf A$ is full row rank ($\operatorname {rank} \mathbf {A}\! =\! N$), $\mathbf T$ is full column rank.
The estimation problem proposed in  Eq. \eqref{Eq:MMYX}, via  Kronecker product,  is turned into a classical LSE formula.

\subsection{Estimation via Neural Network}

 An artificial neural network (ANN) involves a network of simple processing elements (artificial neurons) which can exhibit complex global behavior, determined by the connections between the processing elements and element parameters. In most cases an ANN is an adaptive system that changes its structure based on external or internal information that flows through the network. ANN is the fundamental part of the state-of-the-art  deep neural networks (DNN).
The output and the input of the NN model could be seen as:
\begin{normalsize}
\begin{small}
\[  \mathbf{y}\!=\!\bm f \left( \mathbf{x} \right)\! \triangleq \!{{\bm f}^{L}}\left( {{\mathbf{W}}^{L}}  \!\cdots\! {{\bm f}^{2}}\left( {{\mathbf{W}}^{\text{2}}}{{\bm f}^{1}}\left( {{\mathbf{W}}^{1}}\mathbf{x}\!+\!{{\mathbf{b}}^{1}} \right)\!+\!{{\mathbf{b}}^{\text{2}}} \right)\!\cdots\! \!+\!{{\mathbf{b}}^{L}} \right)\]
\end{small}
\end{normalsize}
or for each layer:
\[\left\{ \begin{aligned}
  & {{\mathbf{z}}^{(l)}}={{\mathbf{W}}^{(l-1)}}{{\mathbf{a}}^{(l-1)}}+{{\mathbf{b}}^{(l-1)}} \\
 & {{\mathbf{a}}^{(l)}}={{\bm f}^{(l)}}\left( {{\mathbf{z}}^{(l)}} \right) \\
\end{aligned} \right.\]
Specifically, $\mathbf{a}^{(1)}$ is the input $\mathbf x$, and $\mathbf{a}^{(L)}$ is the output $\mathbf y$. Note that the parentheses of superscript can be omitted  without creating ambiguity.

For a well-trained NN, the gradient is calculable. Starting form  Eq. \eqref{Eq:MMYX}, there are some connection between the Jacobian Matrix $\mathbf J$ and the gradient.
According to Eq. \eqref{eq:J1}, we face the problem which to solve $\mathbf{J}=\text{d}\frac{\mathbf{y}}{\text{d}{{\mathbf{x}}^{\text{T}}}}$.

With repeated use of the Chain Rule given in Eq. \eqref{eq:pd5},
\begin{equation}
\label{eq:JL}
\begin{aligned}
  \mathbf{J} &=\frac{\partial \mathbf{y}}{\partial {{\mathbf{x}}^{\text{T}}}}\!=\!\frac{\partial {{\mathbf{a}}^{\left( L \right)}}}{\partial {{\mathbf{x}}^{\text{T}}}}\!=\!\frac{\partial {{\bm f}^{\left( L \right)}}\left( {{\mathbf{z}}^{\left( L \right)}} \right)}{\partial {{\mathbf{z}}^{\left( L \right)}}^{\text{T}}}\frac{\partial {{\mathbf{z}}^{\left( L \right)}}}{\partial {{\mathbf{x}}^{\text{T}}}} \\
 & =\!\frac{\partial {{\bm f}^{\left( L \right)}}\left( {{\mathbf{z}}^{\left( L \right)}} \right)}{\partial {{\mathbf{z}}^{\left( L \right)}}^{\text{T}}}\frac{\partial \left( {{\mathbf{W}}^{L-1}}{{\mathbf{a}}^{L-1}}+{{\mathbf{b}}^{L-1}} \right)}{\partial {{\mathbf{x}}^{\text{T}}}} \\
 & =\!\text{diag}\left(\bm f^{ L '}_ {\mathbf z\!=\!{\mathbf{z}}^{ L }}\right) {{\mathbf{W}}^{L-1}}\frac{\partial {{\mathbf{a}}^{L-1}}}{\partial {{\mathbf{x}}^{\text{T}}}}\!  \triangleq \!{{\mathbf{\Lambda }}^{L}}{{\mathbf{W}}^{L-1}}\frac{\partial {{\mathbf{a}}^{L-1}}}{\partial {{\mathbf{x}}^{\text{T}}}}\\
  & ={{\mathbf{\Lambda }}^{L}}{{\mathbf{W}}^{L-1}}{{\mathbf{\Lambda }}^{L-1}}{{\mathbf{W}}^{L-2}}\cdots {{\mathbf{\Lambda }}^{2}}{{\mathbf{W}}^{1}} \\
\end{aligned}
\end{equation}
where ${{\mathbf{\Lambda }}^{l}}\!=\!\text{diag}\left(\bm f^{ l '}_ {\mathbf z\!=\!{\mathbf{z}}^{ l }}\right), l\!=\!2,\cdots,L$

\section{Case Studies}
\label{Sec:Cases}

Cases are built upon the simulation tool MATPOWER \cite{MATPOWER2011matpower}. Given the power injection on each node, i.e. $\mathbf y$, by solving the power flow equation Eq. \eqref{eq:PQend}, the voltage magnitude and angle  are obtained, i.e. , $\mathbf x$ in Eq. \eqref{eq:J1}.

\subsection{IEEE 9-Bus System}
\begin{figure}[htbp]
\centering
\includegraphics[width=0.48\textwidth]{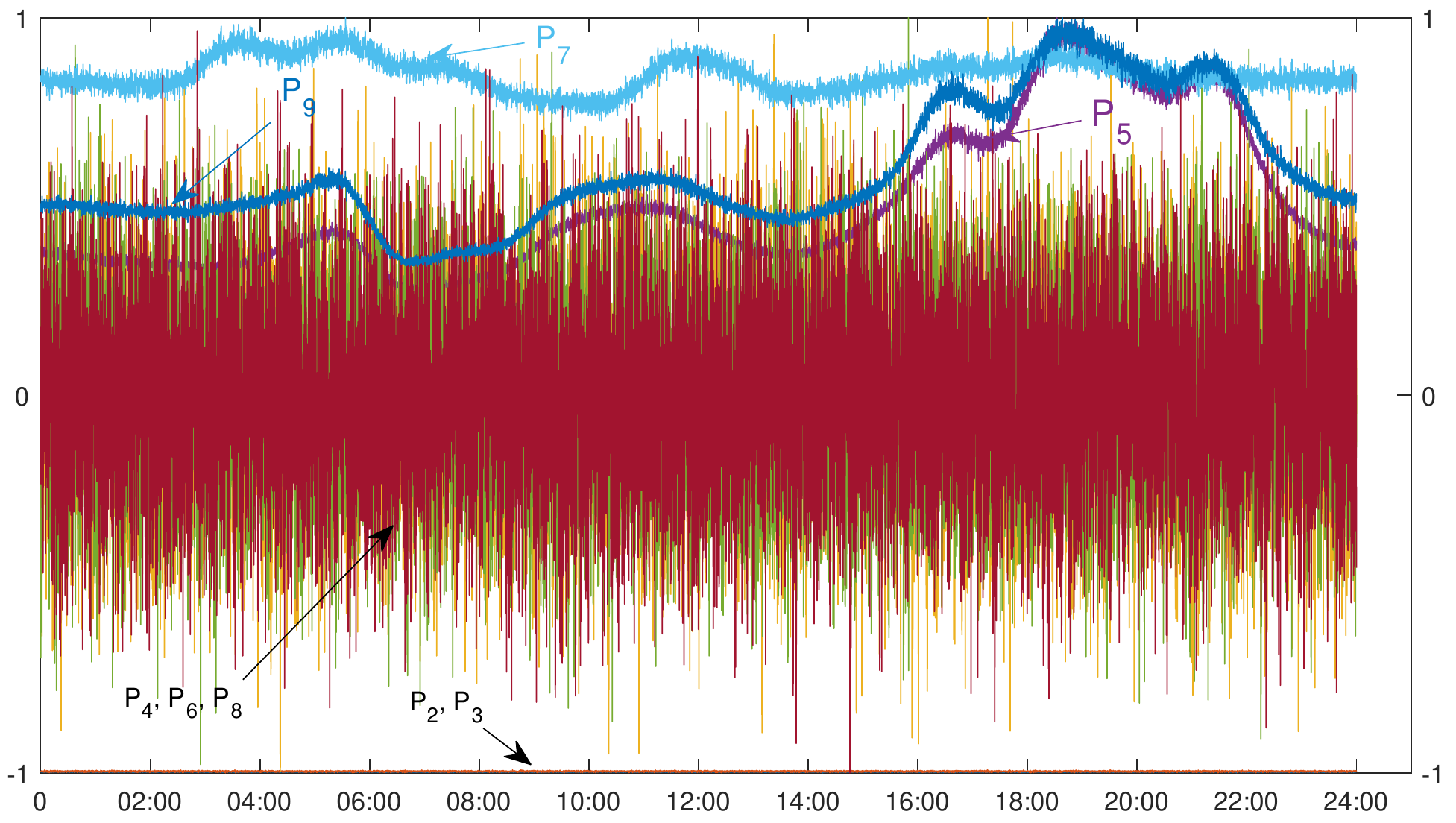}
\caption{Daily Power Consumption for IEEE 9-Bus System}
\label{fig:dailyP}
\end{figure}

For a standard IEEE 9-Bus System, Node 1 is the slack bus, Node 2,3 are the PV buses, Node 5,7,9 are the PQ buses with load injection, and Node 4,6,8 are the PQ buses without load or generator injection, also seemed as tie line. Considering the daily behaviors of power consumption and the power fluctuation according to \cite{he2017invisible}, the normalization value of active power $P$ is obtained as shown in Fig. \ref{fig:dailyP}.  The  reactive power $Q$ has similar trend. Note that the raw value of $P$ and $Q$ of Node 5,7,9 are much larger than that of Node 4,6,8.

\begin{figure}[ht]
\centering
\subfloat[Truth-value of  $\mathbf J_0$]{
\includegraphics[width=0.25\textwidth]{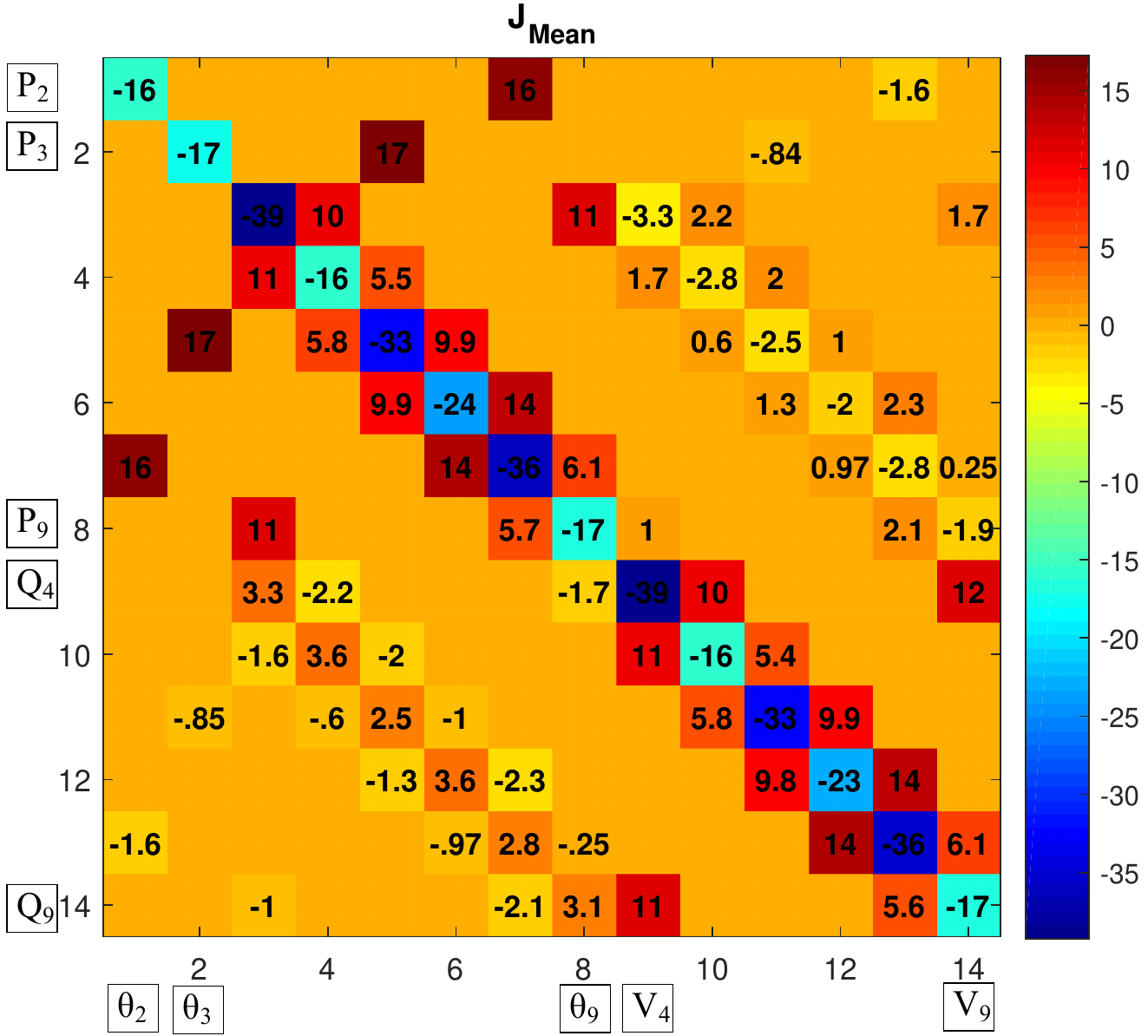}}
\subfloat[Standard deviation of $\mathbf J_0$]{
\includegraphics[width=0.23\textwidth]{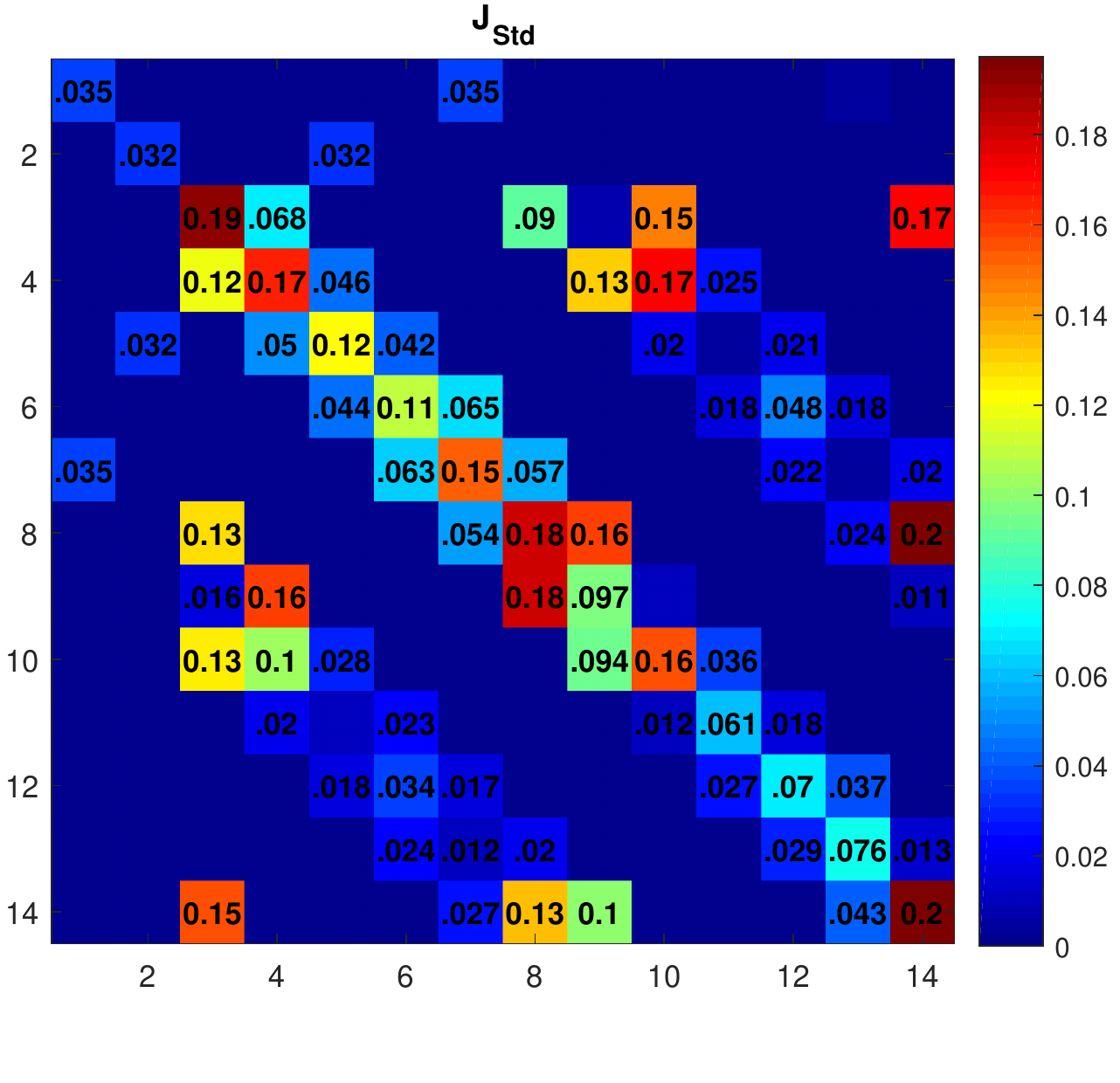}}
\caption{Basic  Statistical Information of $\mathbf J$ from 9600 Samplings}
\label{fig:dailyJ0}
\end{figure}

Suppose the dataset is 9600 points for one day (100 times for 15 minutes).  And then, for each point, our target benchmark/truth-value $\mathbf J_0$ is calculated by the simulator via  Eq. \eqref{eq:HNKLend} in a model-based way. The results are shown in Fig. \ref{fig:dailyJ0}. The variation of the  Jacobian Matrix is so small that we can deduce that  $\mathbf J_0$ is a stable value under the normal daily operation of the network.

\subsubsection{LSE, TLS, WLS without Observation Errors}
{\Text{\\}}

\begin{figure*}[tb]
\centering
\subfloat[LSE: 2400 samplings]{
\includegraphics[width=0.25\textwidth]{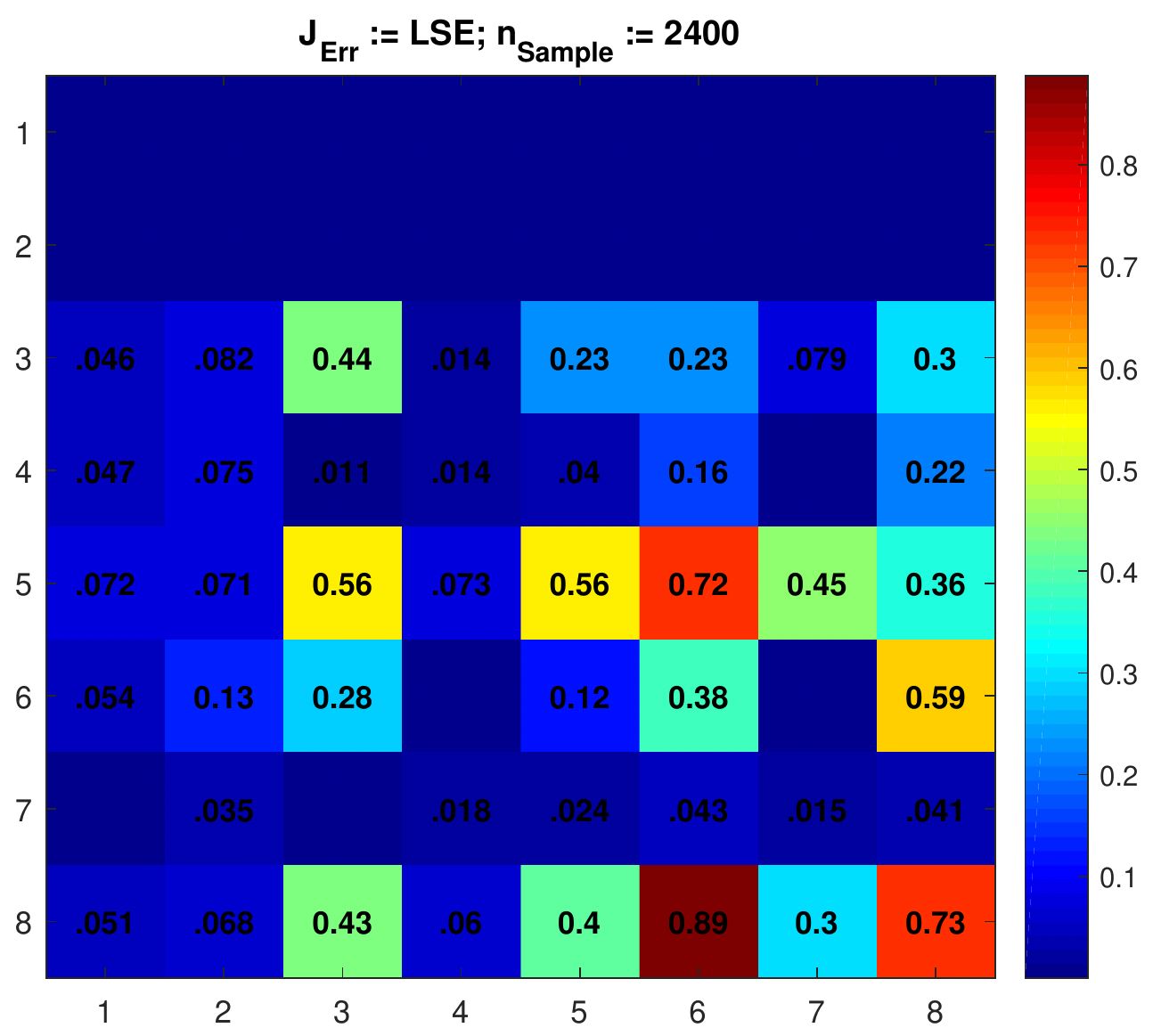}
}
\subfloat[LSE: 240 samplings]{
\includegraphics[width=0.25\textwidth]{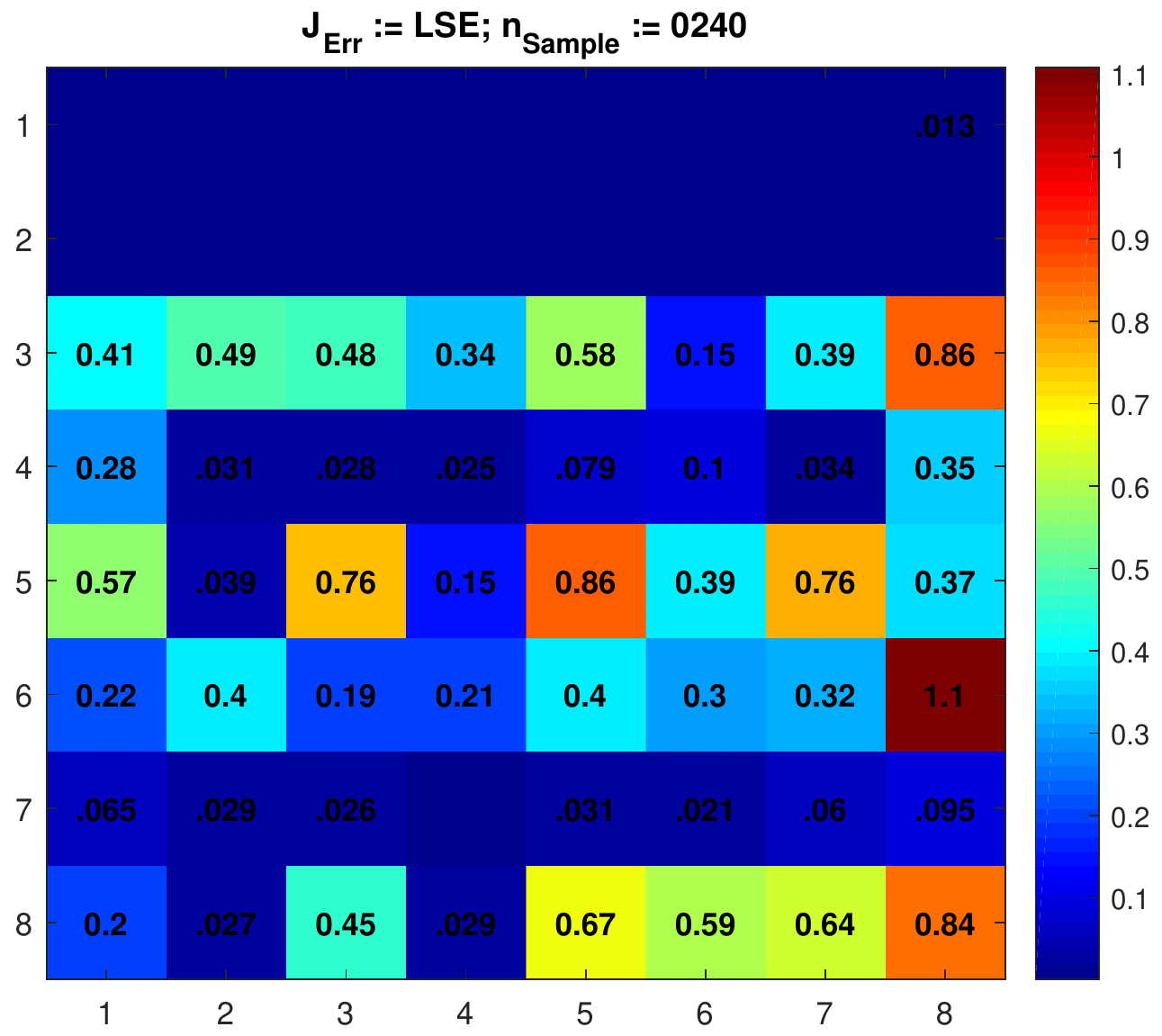}
}
\subfloat[LSE: 96 samplings]{
\includegraphics[width=0.25\textwidth]{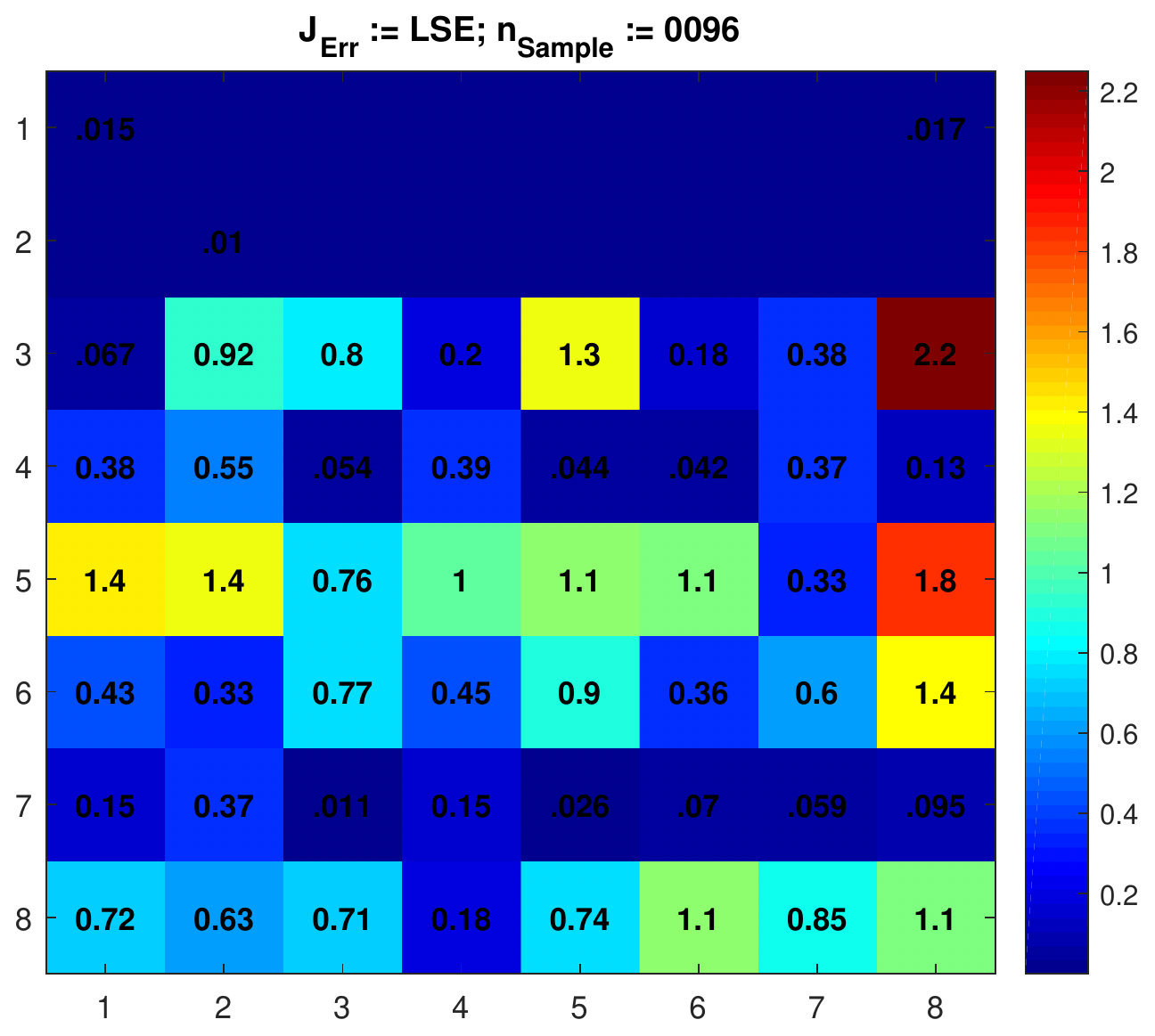}
}
\subfloat[LSE: 5000 samplings, white noises]{
\includegraphics[width=0.25\textwidth]{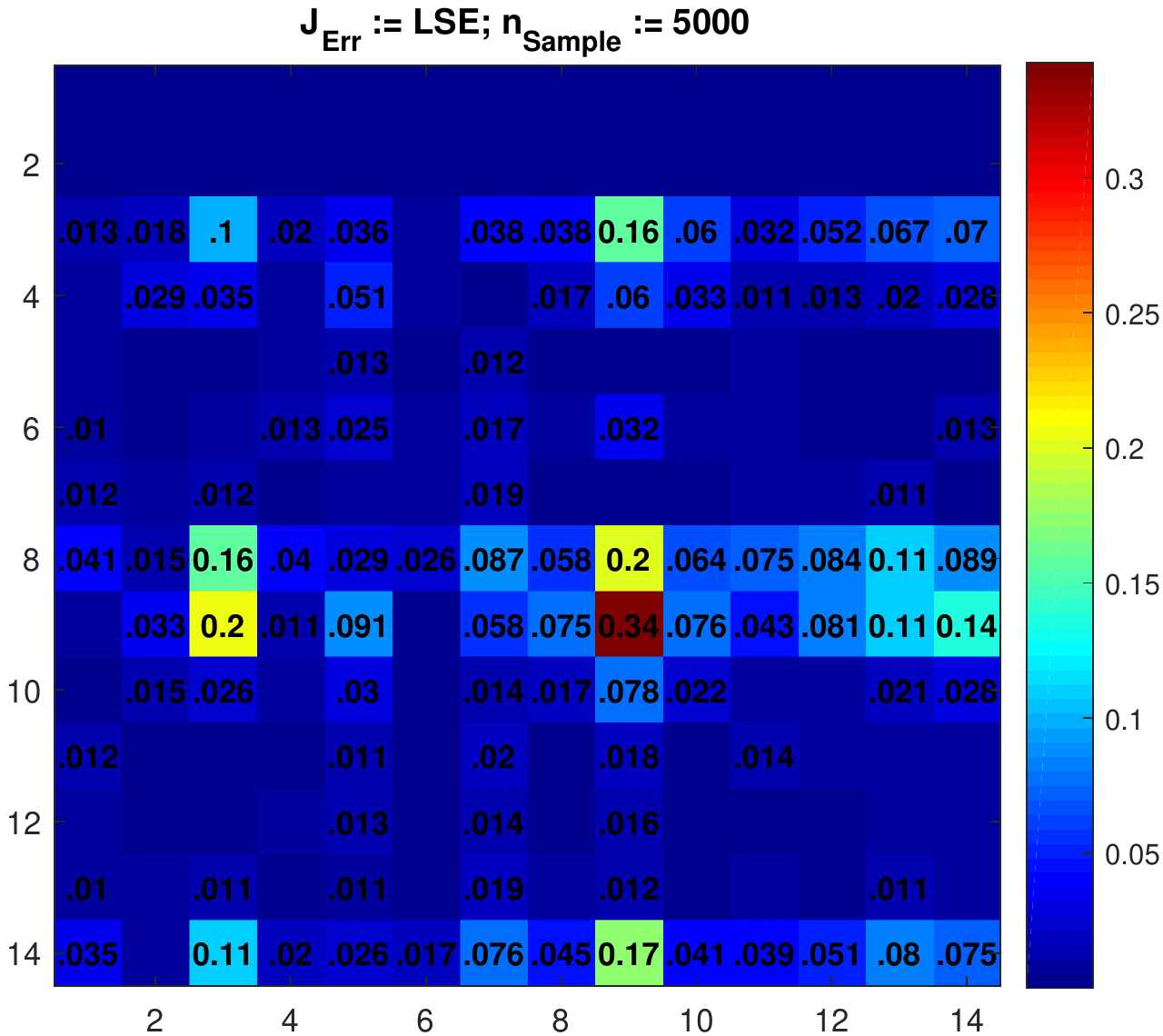}
}

\subfloat[TLS: 2400 samplings]{
\includegraphics[width=0.25\textwidth]{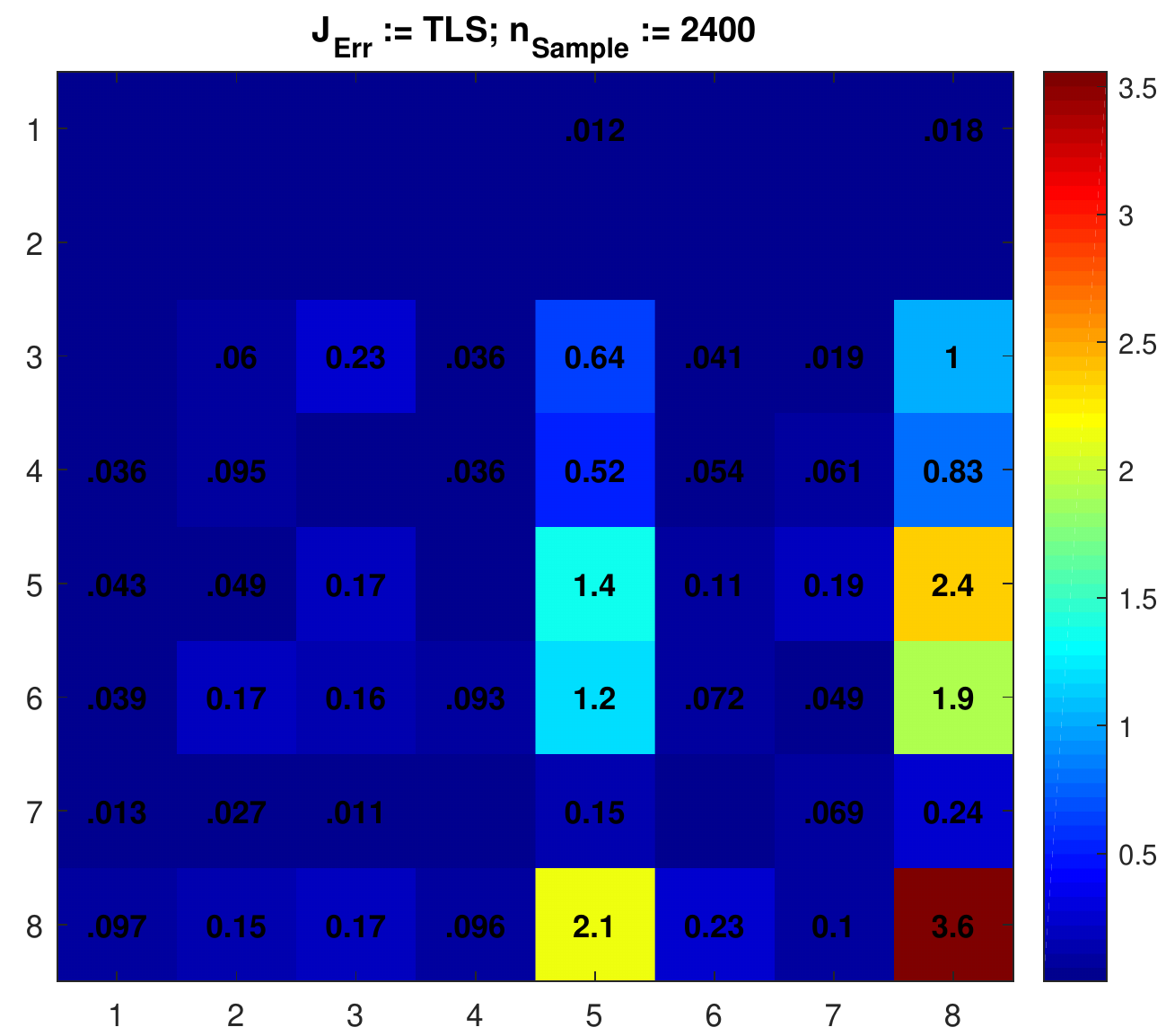}
}
\subfloat[TLS: 240 samplings]{
\includegraphics[width=0.25\textwidth]{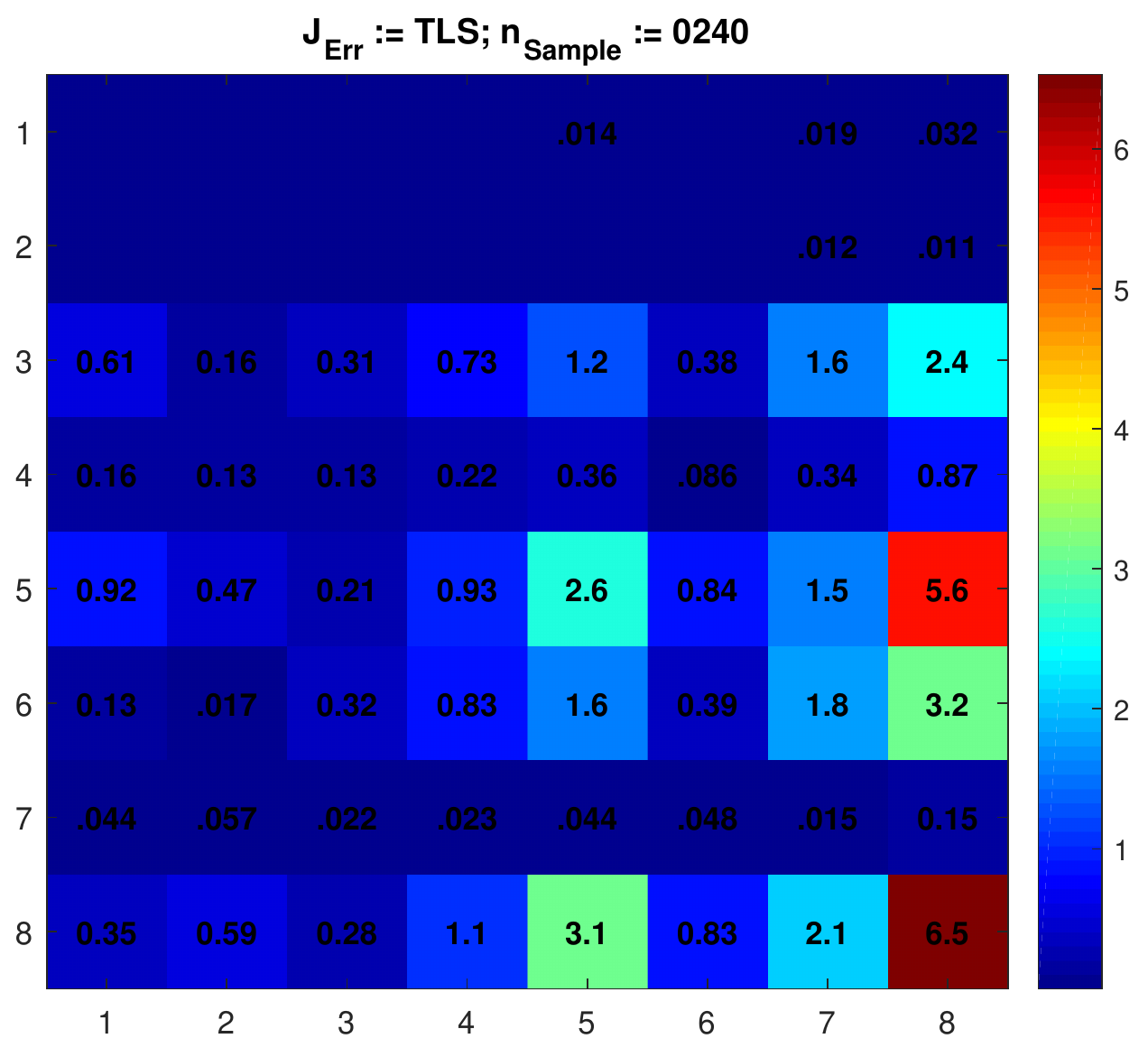}
}
\subfloat[TLS: 96 samplings]{
\includegraphics[width=0.25\textwidth]{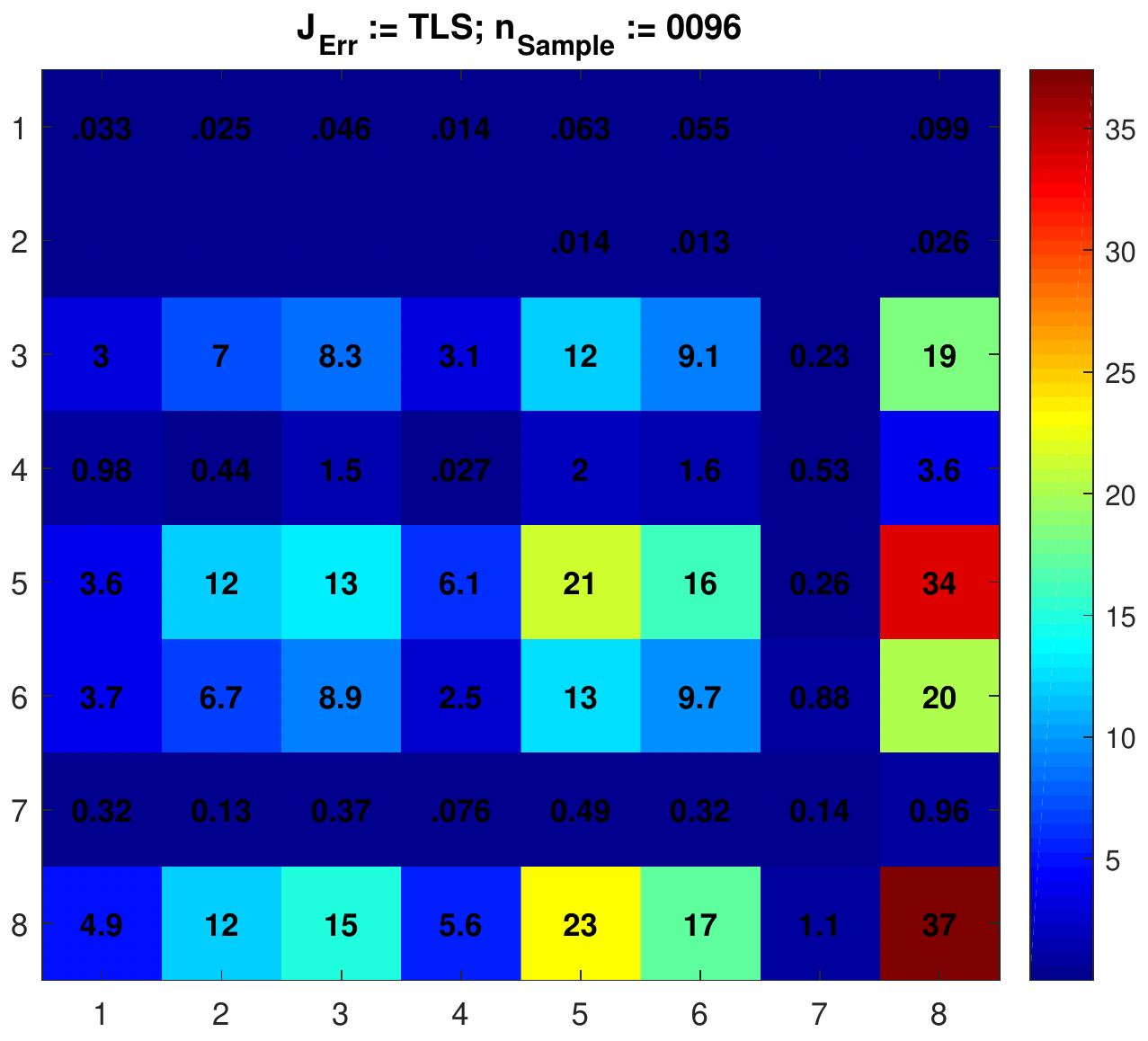}
}
\subfloat[TLS: 5000 samplings, white noises]{
\includegraphics[width=0.25\textwidth]{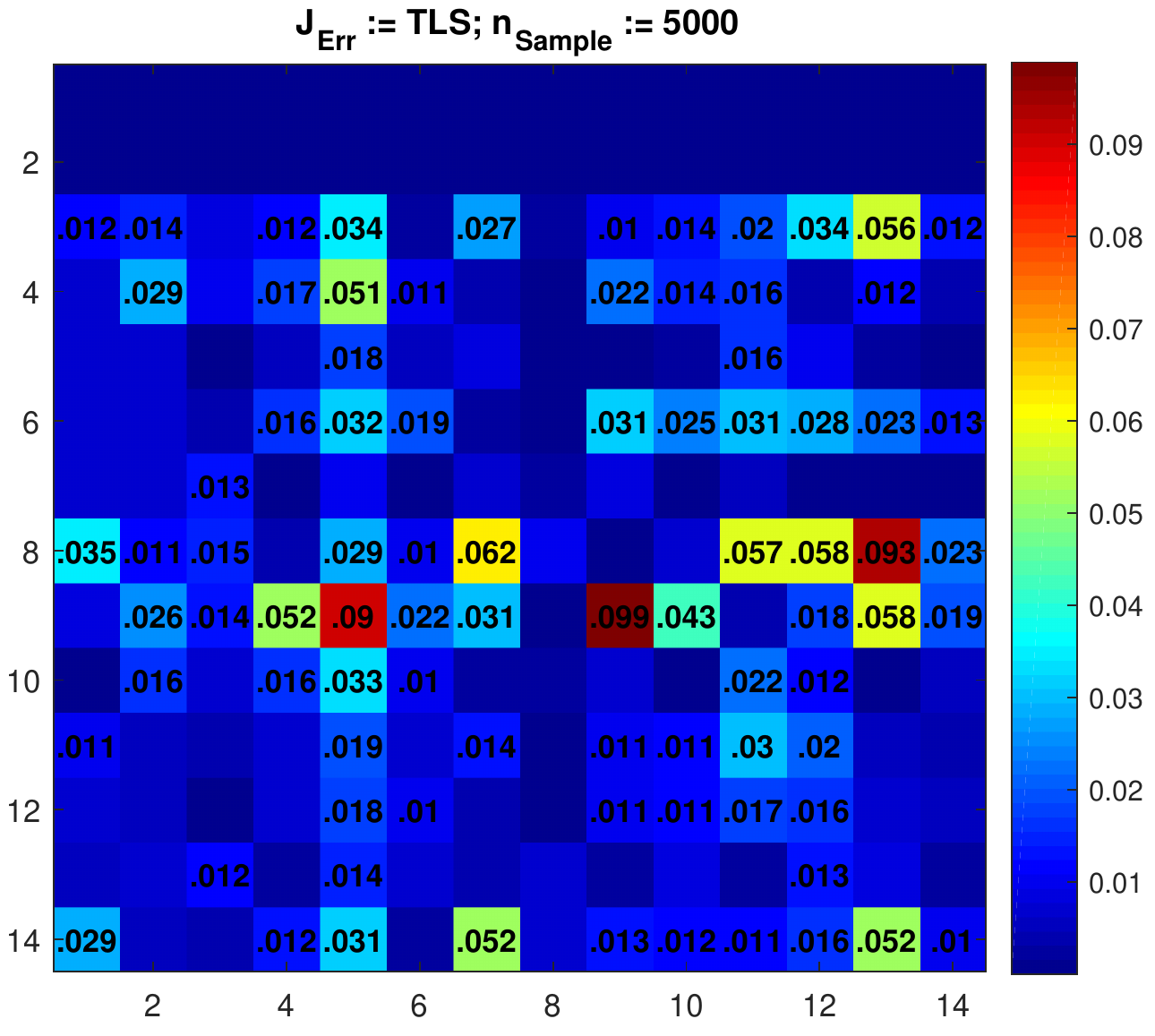}
}

\subfloat[GLS: 2400 samplings]{
\includegraphics[width=0.25\textwidth]{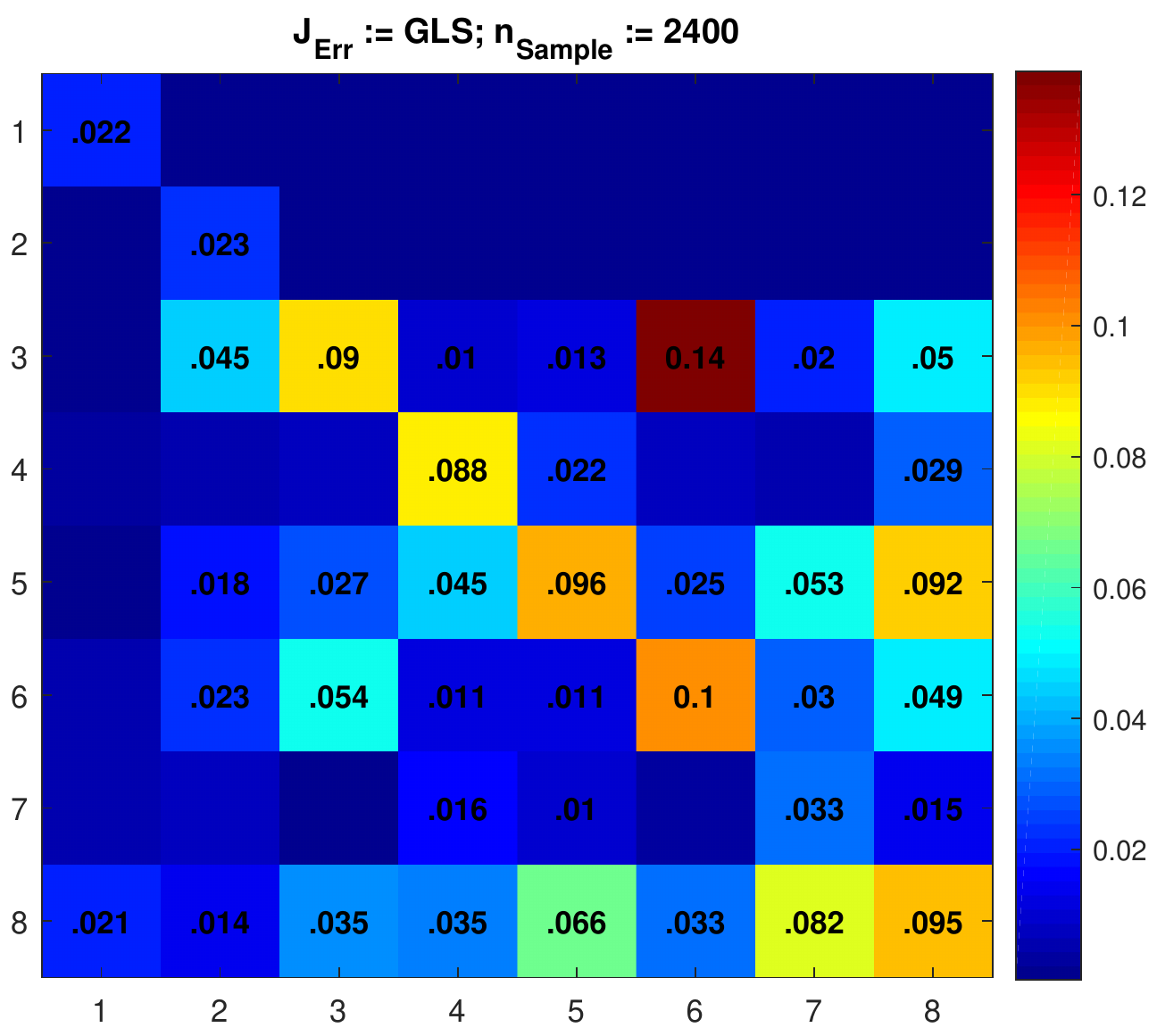}
}
\subfloat[GLS: 240 samplings]{
\includegraphics[width=0.25\textwidth]{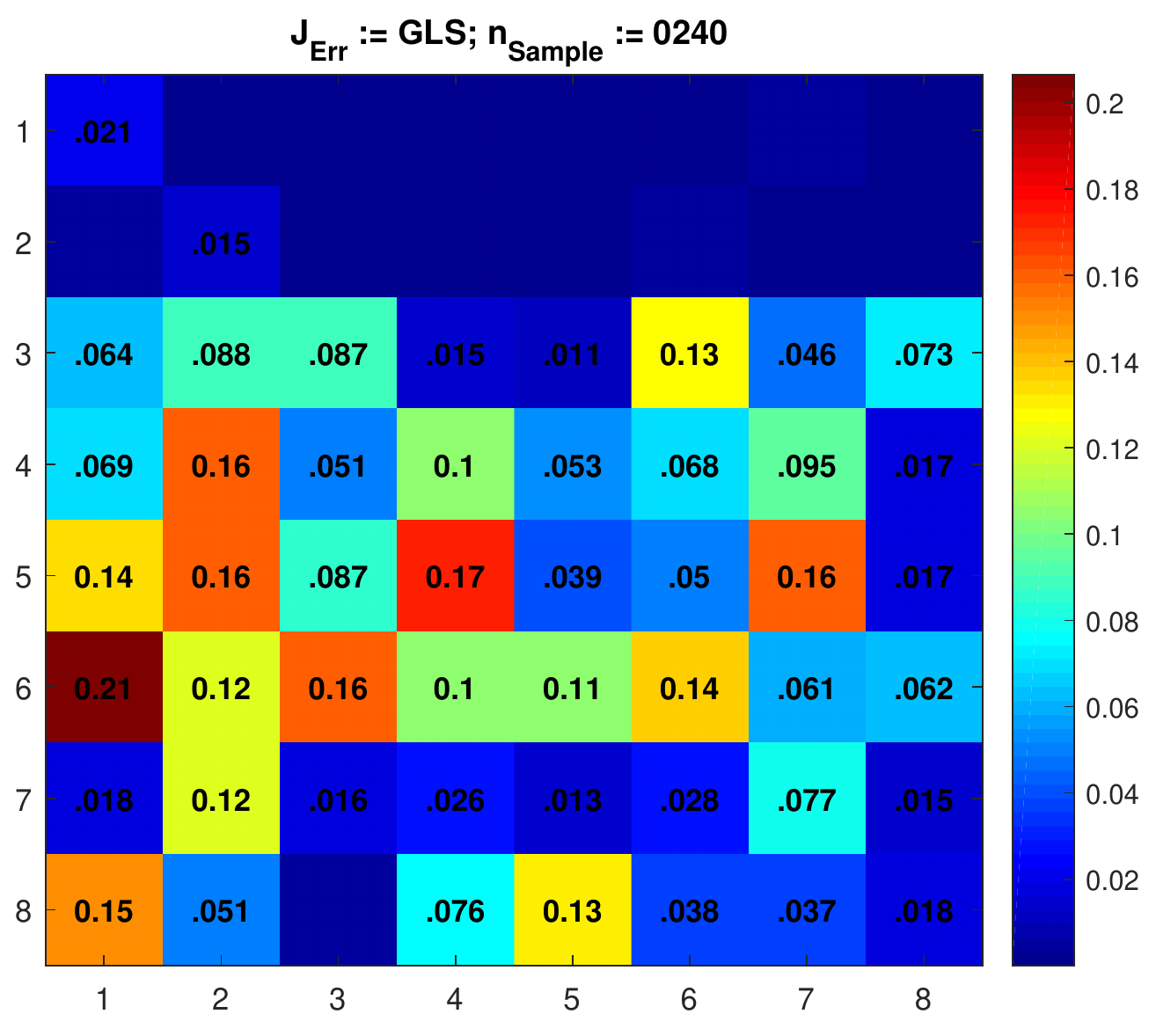}
}
\subfloat[GLS: 96 samplings]{
\includegraphics[width=0.25\textwidth]{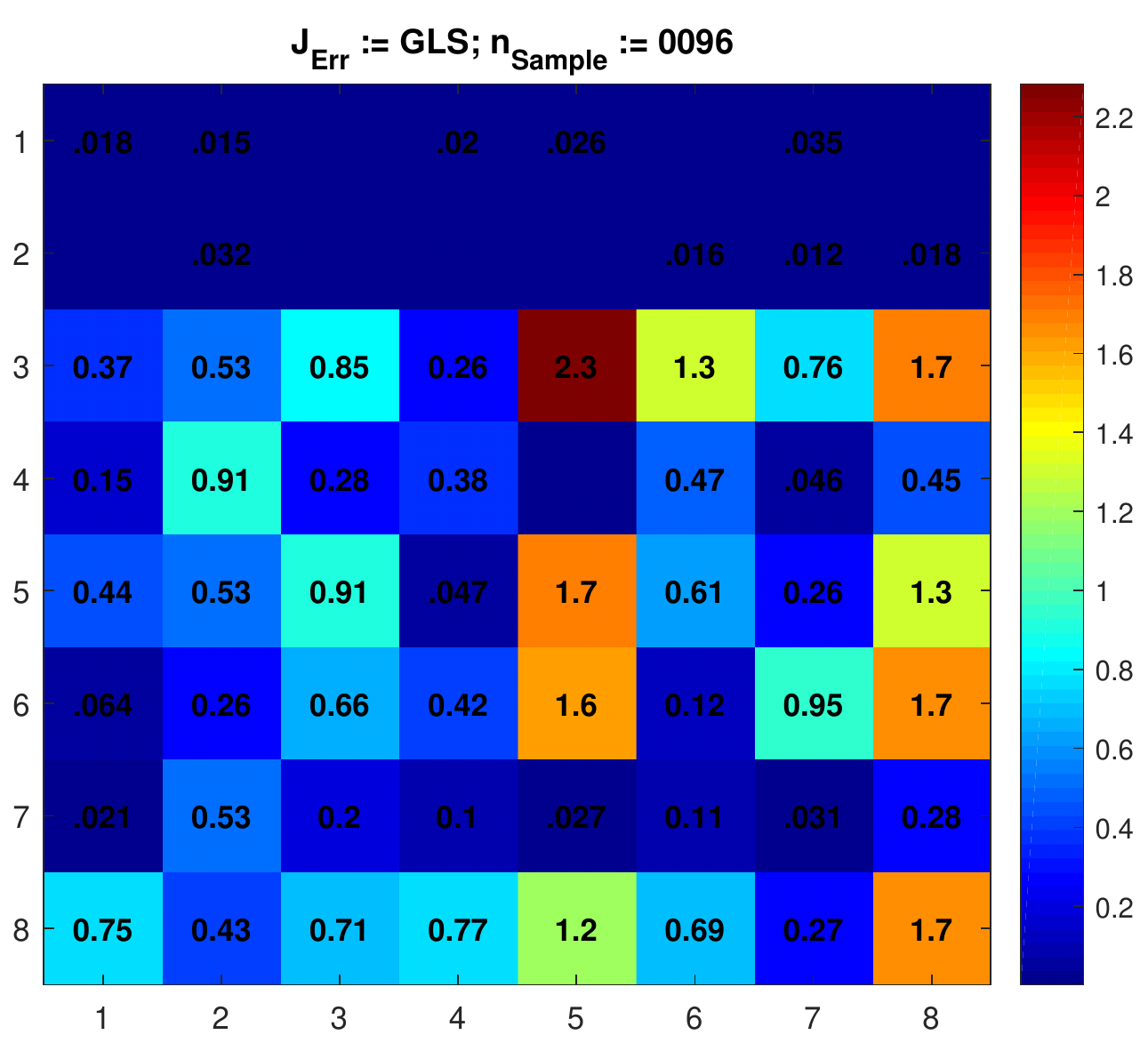}
}
\subfloat[GLS: 5000 samplings, white noises]{
\includegraphics[width=0.25\textwidth]{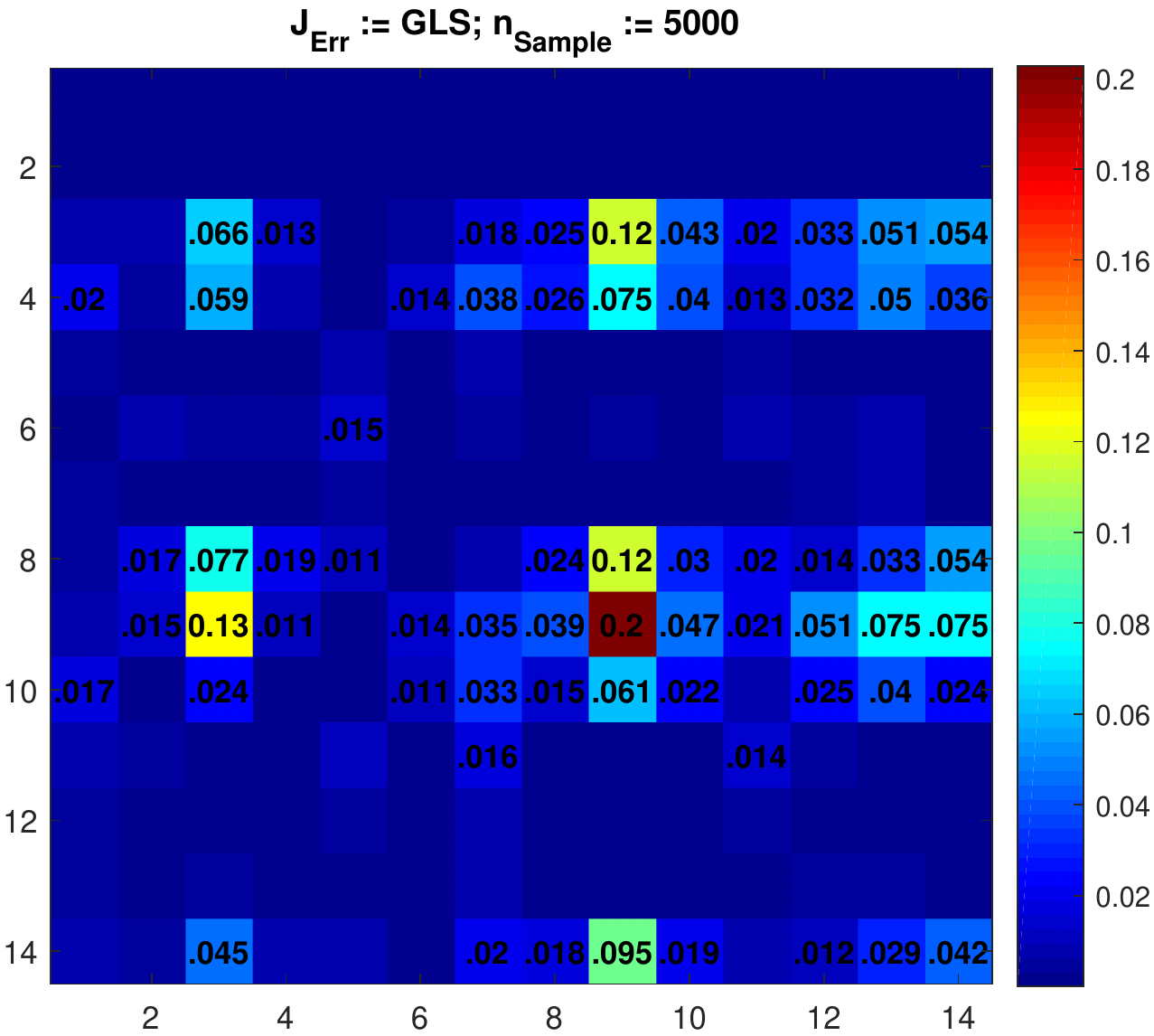}
}
\caption{Estimation Errors of $\mathbf J$ with Large/Medium/Small Dataset: IEEE 9-Bus System without Observation Errors}
\label{fig:J5000}
\end{figure*}

This part assumes that no errors exists  for both $\mathbf x$ and $\mathbf y$. We use the proposed LSE, TLS and WLS algorithm to handle large dataset (2400 points, 100 samplings per hour), medium dataset (240 points, 10 samplings/h), and small dataset (96 points, 4 samplings/h), respectively. We also design a case which simulates the data at a fast sampling rate (e.g. 30 Hz for PMU) within a short time, during which the white noise plays a dominant part of the signals. And we assume there are 5000 samplings for this condition.
The result of the above four scenes are shown in Fig. \ref{fig:J5000}.

Fig. \ref{fig:J5000} shows that the LSE, TLS and GLS all have good performances on the estimation of Jacobian Matrix $\mathbf J$ using a large dataset.  And among them, GLS performs best. For the white noise scenes, however, TLS performs best. It is worth  mentioning that, for GLS,  the calculation of inverse $\mathbf G$ (invG = pinv(G) for GLS Octave/Matlab code) takes a lot of computing resources.
With a medium dataset, the performance of LSE, TLS and WLS  become worse. The GLS performs much better than LSE,  and the TLS performs worst.
With a small dataset,  the TLS result is no longer acceptable, and there are no obvious difference between LSE and GLS.
Note that WLS is a special case of GLS.
We deduce that \textbf{the GLS performs better as the dataset increase, and TLS may be suitable to handle white noises.}

\subsubsection{LSE, TLS, WLS with errors from observation}
{\Text{\\}}

\begin{figure}[ht]
\centering
\subfloat[Error from $\mathbf y$: GLS]{\label{fig:J200EY}
\includegraphics[width=0.25\textwidth]{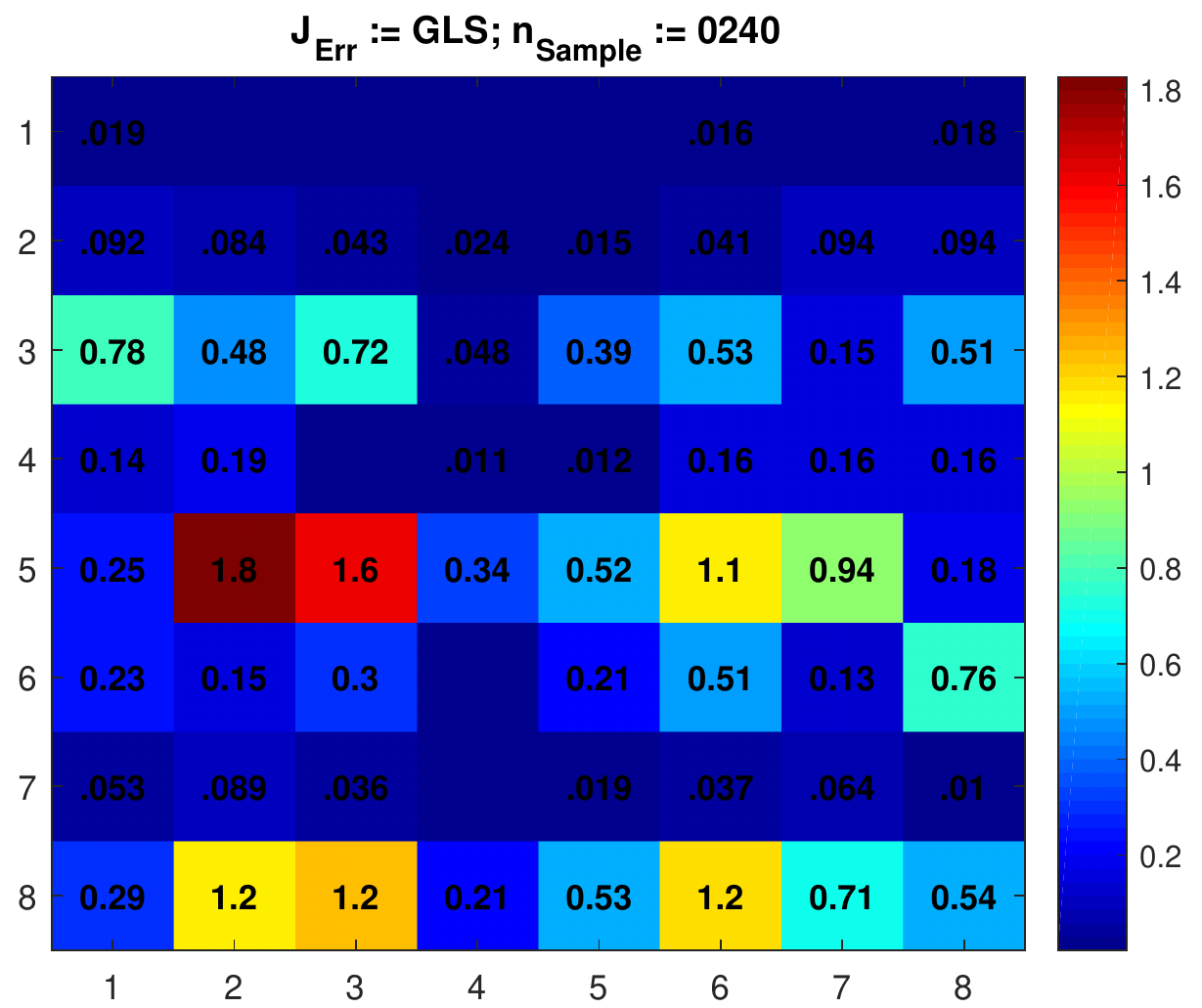}
}
\subfloat[Error from $\mathbf x$: TLS]{\label{fig:J200EX}
\includegraphics[width=0.24\textwidth]{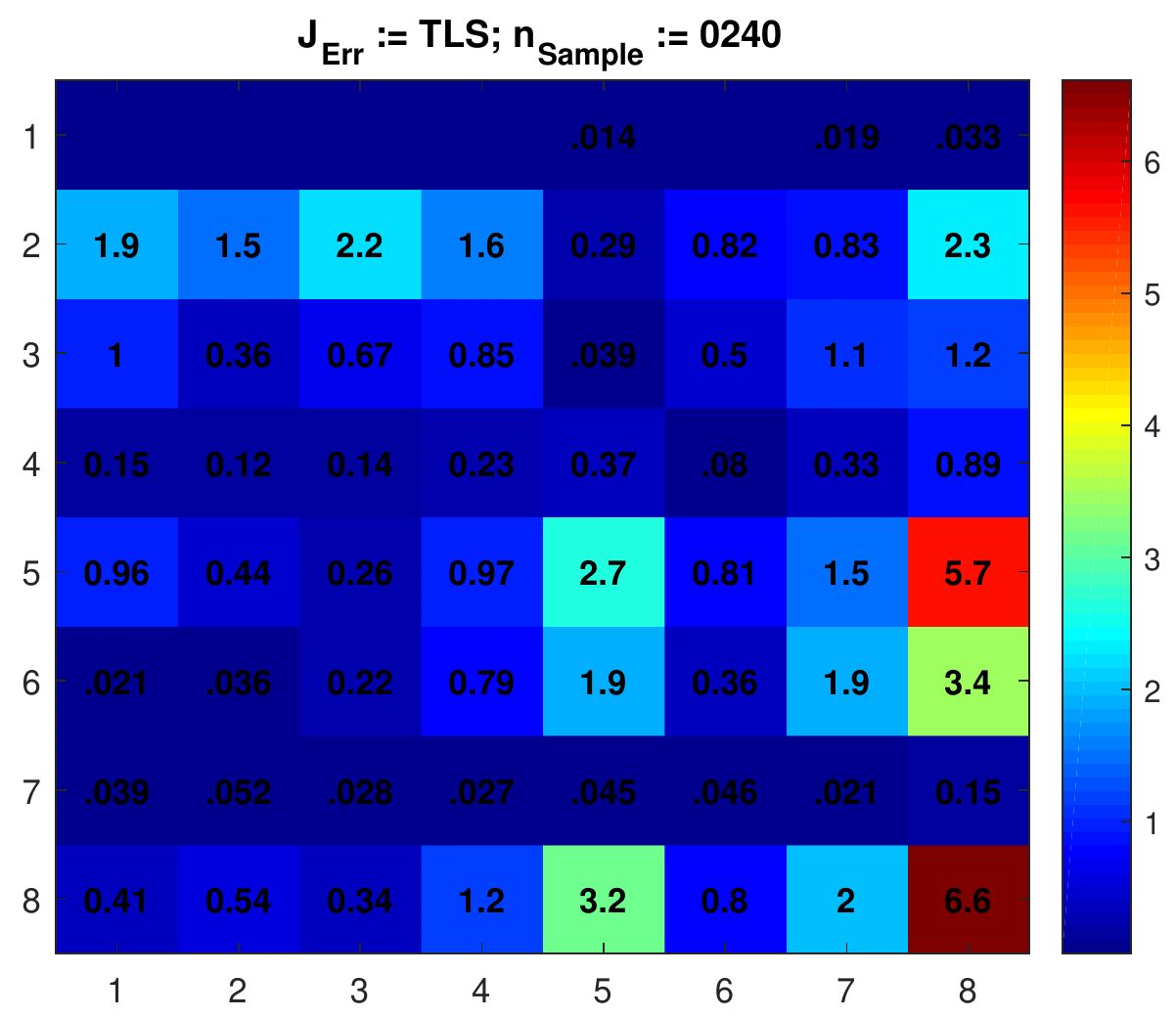}
}
\caption{Estimation Errors of $\mathbf J$ Considering Observation Errors}
\label{fig:J200Y}
\end{figure}

This part assumes that there exist errors from the observation variable, active and reactive power of each node, i.e.,  $\mathbf y$.
Suppose that
\[\hat y(n,t) := y_0(n,t)+\alpha_1 \cdot{} r_1 +\alpha_2\]
where $n$ is the $n$-th node, t is the sampling time,
and $y_0$ is the truth-value, $\alpha_1 \cdot{} r_1$ is the variance of the errors and $\alpha_2$ is the bias.
$r_1$ is a standard Gaussian noise and  $\alpha_1, \alpha_2$ are small coefficient relevant to Signal-Noise Rate.
The existence of $\alpha_2$ would not affect the data-driven results because that it does not change $\Delta \mathbf y$. It means that \textbf{the data-driven estimation is robust against fixed measurement errors.} We assume medium dataset are used for this case, the fluctuation of $\mathbf y$  for Node 3 during sampling time $t_s=[60:100]$  is $4+5\text{i}$ MVA , for Node 5 during  $t_s=[50:90]$ is $3+2\text{i}$ MVA.
Running the code, we find that only GLS result is acceptable, as shown in Fig~\ref{fig:J200EY}. \textbf{The GLS is adapted to the scene that the observed data is not reliable during some part of the cycle.}

On the other side, if the error comes from voltage angle and the error variance keeps the whole observation. The TLS performs best, as shown in Fig~\ref{fig:J200EX}.   \textbf{The TLS is adapted to the scene that the  variance comes from  $\mathbf x$ during the whole observation.}

\subsubsection{Neural Network}
{\Text{\\}}

\begin{figure}[ht]
\centering
\subfloat[Prediction of P on load nodes]{\label{fig:NN1}
\includegraphics[width=0.24\textwidth]{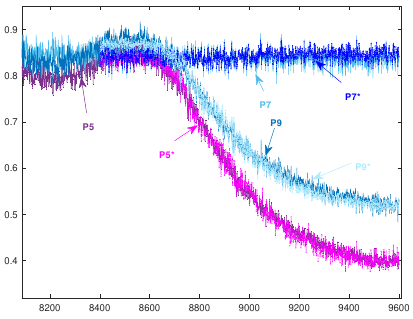}
}
\subfloat[$\mathbf J$ estimation via Eq. \eqref{eq:JL}]{\label{fig:NN2}
\includegraphics[width=0.22\textwidth]{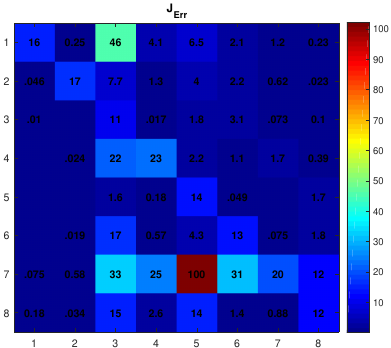}
}
\caption{Power Prediction and $\mathbf J$ Estimation using NN}
\label{fig:NN}
\end{figure}

This section explores the state-of-the-art  deep neural networks (DNN). We use a 5 layers NN, and choose tanh as the activate function. The neural number for each layers is $[14, 50, 50, 50, 14]$. Taking $\mathbf x$ in Eq. \eqref{eq:J1} as the input of the NN, and $\mathbf y$ as the output. We use the data during 1:8400 (00:00--21:00) for training, and during 8401:9600 (21:00--24:00) for testing. Taking the PQ buses with load injection (i.e., Node 5,7 and 9), the truth-value $P_5, P_7, P_9$ and predicted value $P_5^*, P_7^*, P_9^*$ of active power P are shown in Fig. \ref{fig:NN1}. The NN performs pretty well on the prediction task. We then use Eq. \eqref{eq:JL} to calculate Jacobian Matrix, and  find that the $\mathbf J$ estimation task fails, as shown in Fig. \ref{fig:NN2}.
It can deduce that \textbf{the direct use of NN may be not suitable to handle derivative signal analysis. } The derivative signal may have some connection with the residential network, and this topic will be discussed elsewhere.

\section{Conclusion}
\label{section: concl}
\normalsize{}
This paper explores a data-driven method for Jacobian Matrix estimation. The least-squared errors (LSE) algorithm and its variants total least-squared (TLS) and generalized least-squared (GLS), as well as neural network (NN) based estimation algorithm are studied.
These algorithms are data-driven and sensitive to up-to-date topology parameters and state variables.
The data-driven estimation is also robust against fixed measurement errors.

The theories, equations, derivations, codes, advantages and disadvantages, and application scenes of each algorithm are discussed. The GLS is adapted to the scene that the observed data is not reliable during some part of the cycle, and performs better as the dataset increase. The TLS  is adapted to the scene that the error variance comes from  $\mathbf x$ during the whole observation, and TLS may be used to handle white noises. Directly use of NN is not suitable to handle derivative signal analysis, the derivative signal may have some connection with the state-of-the-art residential network.
\appendices

%

\section{Active and Reactive Power for Each Node}
\label{Sec:PQProc}
Starting with Eq. \eqref{eqSS}, Eq. \eqref{eq:PQend} is obtained. The procedures are as follows:
\begin{normalsize}
\begin{small}
\begin{equation}
\label{eq:PQProc}
\begin{aligned}
  {{S}_{i}}& ={{P}_{i}}+\text{j}{{Q}_{i}}={{{\dot{U}}}_{i}}\sum\limits_{k}{\overline{{{y}_{ik}}}\overline{{{{\dot{U}}}_{k}}}}={{{\dot{U}}}_{i}}\sum\limits_{k\ne i}{\overline{{{Y}_{ik}}}\overline{{{{\dot{U}}}_{k}}}}+{{{\dot{U}}}_{i}}\overline{{{Y}_{ii}}}\overline{{{{\dot{U}}}_{i}}} \\
 &     ={{V}_{i}}{{\text{e}}^{\text{j}{{\theta }_{i}}}}\sum\limits_{k\ne i}{\left( {{G}_{ik}}\!-\!\text{j}{{B}_{ik}} \right){{V}_{k}}{{\text{e}}^{-\text{j}{{\theta }_{k}}}}}+{{V}_{i}}^2\left( \!-\sum\limits_{k\ne i}{\left( {{G}_{ik}}\!-\!\text{j} {{B}_{ik}} \right)} \right) \\
 & ={{V}_{i}}\sum\limits_{k\ne i}{{{V}_{k}}\left( {{G}_{ik}}\text{cos}{{\theta }_{ik}}\!+\!{{B}_{ik}}\text{sin}{{\theta }_{ik}} \right)\!-\!{{V}_{i}}^{2}\sum\limits_{k\ne i}{{{G}_{ik}}}} \\
 &\quad{}+\!\text{j}\left[ {{V}_{i}}\sum\limits_{k\ne i}{{{V}_{k}}\left( {{G}_{ik}}\sin {{\theta }_{ik}}-{{B}_{ik}}\text{cos}{{\theta }_{ik}} \right)}\!+\!{{V}_{i}}^{2}\sum\limits_{k\ne i}{{{B}_{ik}}} \right] \\
 \end{aligned}
\end{equation}
\end{small}
\end{normalsize}

Thus,
\begin{normalsize}
\begin{small}
\[
\begin{aligned}
  & \left\{ \begin{aligned}
  & {{P}_{i}}={{V}_{i}}\sum\limits_{k\ne i}{{{V}_{k}}\left( {{G}_{ik}}\text{cos}{{\theta }_{ik}}+{{B}_{ik}}\text{sin}{{\theta }_{ik}} \right)}-{{V}_{i}}^{2}\sum\limits_{k\ne i}{{{G}_{ik}}} \\
 & {{Q}_{i}}={{V}_{i}}\sum\limits_{k\ne i}{{{V}_{k}}\left( {{G}_{ik}}\sin {{\theta }_{ik}}-{{B}_{ik}}\text{cos}{{\theta }_{ik}} \right)}+{{V}_{i}}^{2}\sum\limits_{k\ne i}{{{B}_{ik}}} \\
\end{aligned} \right. \\
 &      \\
\end{aligned}
\]
\end{small}
\end{normalsize}

\section{Jacobian Matrix and its Four Components}
\label{Sec:HNKLProc}
Starting with Eq. \eqref{eq:PQend}, Eq. \eqref{eq:HNKLend} is obtained. The procedures are as follows.
Take ${H}_{ij}$ for instance, and similar to ${N}_{ij}$,${K}_{ij}$,${L}_{ij}$:\\
if $ i\ne j $, $ {{H}_{ij}}=\frac{\partial {{P}_{i}}}{\partial {{\theta }_{j}}}={{V}_{i}}{{V}_{j}}\left( {{G}_{ij}}\text{sin}{{\theta }_{ij}}-{{B}_{ij}}\text{cos}{{\theta }_{ij}} \right)$; \\
if $ i=j$,
${{H}_{ii}}={{V}_{i}}\sum\limits_{k\ne i}{{{V}_{k}}\left( \!-\!{{G}_{ik}}\text{sin}{{\theta }_{ik}}\!+\!{{B}_{ik}}\text{cos}{{\theta }_{ik}} \right)}  \!\\ \qquad{}\qquad{}\qquad{}\quad{}=-\!{{Q}_{i}}\!+\!{{V}_{i}}^{2}\sum\limits_{k\ne i}{{{B}_{ik}}}\!+\!{{V}_{i}}^{2}{{b}_{i}}\!=\!\!-\!{{V}_{i}}^{2}{{B}_{ii}}\!-\!{{Q}_{i}}\!+\!{{V}_{i}}^{2}{{b}_{i}}$.

\section{SVD for augmented Matrix}
\label{Sec:SVDXYZ}
It is known that:
\[\left\{ \begin{aligned}
  & \mathbf{X}={{\mathbf{U}}_{\mathbf{X}}}{{\mathbf{\Sigma }}_{\mathbf{X}}}{{\mathbf{V}}_{\mathbf{X}}}^{\text{T}} \\
 & \mathbf{Y}={{\mathbf{U}}_{\mathbf{Y}}}{{\mathbf{\Sigma }}_{\mathbf{Y}}}{{\mathbf{V}}_{\mathbf{Y}}}^{\text{T}} \\
\end{aligned} \right.\Rightarrow \mathbf{Z}=\left[ \begin{matrix}
   \mathbf{X} & \mathbf{Y}  \\
\end{matrix} \right]={{\mathbf{U}}_{\mathbf{Z}}}{{\mathbf{\Sigma }}_{\mathbf{Z}}}{{\mathbf{V}}_{\mathbf{Z}}}^{\text{T}}\]

Giving
\[\begin{aligned}
   \mathbf{Z}{{\mathbf{Z}}^{\text{T}}}&=\left[ \begin{matrix}
   \mathbf{X} & \mathbf{Y}  \\
\end{matrix} \right]\left[ \begin{matrix}
   {{\mathbf{X}}^{\text{T}}}  \\
   {{\mathbf{Y}}^{\text{T}}}  \\
\end{matrix} \right]=\mathbf{X}{{\mathbf{X}}^{\text{T}}}+\mathbf{Y}{{\mathbf{Y}}^{\text{T}}} \\
 & ={{\mathbf{U}}_{\mathbf{X}}}{{\mathbf{\Sigma }}_{\mathbf{X}}}^{\text{2}}{{\mathbf{U}}_{\mathbf{X}}}^{\text{T}}\text{+}{{\mathbf{U}}_{\mathbf{Y}}}{{\mathbf{\Sigma }}_{\mathbf{Y}}}^{\text{2}}{{\mathbf{U}}_{\mathbf{Y}}}^{\text{T}} \\
 & =\left[ \begin{matrix}
   {{\mathbf{U}}_{\mathbf{X}}} & {{\mathbf{U}}_{\mathbf{Y}}}  \\
\end{matrix} \right]\left[ \begin{matrix}
   {{\mathbf{\Sigma }}_{\mathbf{X}}}^{\text{2}} & \mathbf{0}  \\
   \mathbf{0} & {{\mathbf{\Sigma }}_{\mathbf{Y}}}^{\text{2}}  \\
\end{matrix} \right]\left[ \begin{matrix}
   {{\mathbf{U}}_{\mathbf{X}}}^{\text{T}}  \\
   {{\mathbf{U}}_{\mathbf{Y}}}^{\text{T}}  \\
\end{matrix} \right]
\end{aligned}\]
so that
\[\left\{ \begin{aligned}
  & {{\mathbf{U}}_{\mathbf{Z}}}=\left[ \begin{matrix}
   {{\mathbf{U}}_{\mathbf{X}}} & {{\mathbf{U}}_{\mathbf{Y}}}  \\
\end{matrix} \right] \\
 & {{\mathbf{\Sigma }}_{\mathbf{Z}}}=\left[ \begin{matrix}
   {{\mathbf{\Sigma }}_{\mathbf{X}}} & \mathbf{0}  \\
   \mathbf{0} & {{\mathbf{\Sigma }}_{\mathbf{Y}}}  \\
\end{matrix} \right] \\
\end{aligned} \right.\]

\section{Matrix Operation and  Least Squares Method}
\label{Sec:MatrixOpr}
\subsection{Partial Differentia in Matrix/Vector Form}
\begin{equation}
\label{eq:pd1}
\begin{aligned}
   \frac{\partial {{\mathbf{x}}^{\text{T}}}\mathbf{Wx}}{\partial \mathbf{x}}&=\frac{\partial \sum{{{x}_{i}}{{\mathbf{E}}_{1i}}{{w}_{jk}}{{\mathbf{E}}_{jk}}{{x}_{l}}{{\mathbf{E}}_{l1}}}}{\partial \mathbf{x}}=\sum{\frac{\partial {{w}_{il}}{{x}_{i}}{{x}_{l}}}{\partial {{x}_{k}}}{{\mathbf{E}}_{k}}} \\
 & =\sum{\left( {{w}_{kl}}{{x}_{l}}+{{w}_{ik}}{{x}_{i}} \right){{\mathbf{E}}_{k}}}=\left( \mathbf{W}+{{\mathbf{W}}^{\text{T}}} \right)\mathbf{x}
\end{aligned}
\end{equation}
because that
\[\begin{aligned}
  \mathbf{Wx}& =\sum{{{w}_{ij}}{{x}_{k}}{{\mathbf{E}}_{ij}}{{\mathbf{E}}_{k}}}=\sum{{{w}_{ij}}{{x}_{j}}{{\mathbf{E}}_{i}}} \\
  {{\mathbf{W}}^{\text{T}}}\mathbf{x}&=\sum{{{w}_{ij}}{{x}_{k}}{{\mathbf{E}}_{ji}}{{\mathbf{E}}_{k}}}=\sum{{{w}_{ij}}{{x}_{i}}{{\mathbf{E}}_{j}}}
\end{aligned}\]

\begin{equation}
\label{eq:pd2}
\frac{\partial {{\mathbf{y}}^{\text{T}}}}{\partial \mathbf{x}}=\left[ \begin{matrix}
   \frac{\partial {{y}_{1}}}{\partial \mathbf{x}} & \frac{\partial {{y}_{2}}}{\partial \mathbf{x}} & \cdots   \\
\end{matrix} \right]
\end{equation}
for a special case $\mathbf{y}=\mathbf{x}$: $
\frac{\partial {{\mathbf{x}}^{\text{T}}}}{\partial \mathbf{x}}=\mathbf{I}$

\begin{equation}
\label{eq:pd3}
\begin{aligned}
  \frac{\partial {{\mathbf{y}}^{\text{T}}}\mathbf{Wy}}{\partial \mathbf{x}}& =\sum{\frac{\partial {{\mathbf{y}}^{\text{T}}}\mathbf{Wy}}{\partial {{y}_{i}}}\frac{\partial {{y}_{i}}}{\partial \mathbf{x}}}\!
  =\!\left[ \begin{matrix}
   \frac{\partial {{y}_{1}}}{\partial \mathbf{x}} & \frac{\partial {{y}_{2}}}{\partial \mathbf{x}} & \cdots   \\
\end{matrix} \right]\frac{\partial {{\mathbf{y}}^{\text{T}}}\mathbf{Wy}}{\partial \mathbf{y}} \\
 & =\frac{\partial {{\mathbf{y}}^{\text{T}}}}{\partial \mathbf{x}}\frac{\partial {{\mathbf{y}}^{\text{T}}}\mathbf{Wy}}{\partial \mathbf{y}}
\end{aligned}
\end{equation}

\subsection{Chain Rule}
\begin{equation}
\label{eq:pd5}
\frac{\partial \mathbf{y}}{\partial {{\mathbf{x}}^{\text{T}}}}\triangleq \frac{\partial \bm f\left( \mathbf{z} \right)}{\partial {{\mathbf{x}}^{\text{T}}}}=\frac{\partial \bm f\left( \mathbf{z} \right)}{\partial {{\mathbf{z}}^{\text{T}}}}\frac{\partial \mathbf{z}}{\partial {{\mathbf{x}}^{\text{T}}}}
\end{equation}
\textbf{Proof:}\\
Suppose that $\mathbf{C}=\frac{\partial \mathbf{y}}{\partial {{\mathbf{x}}^{\text{T}}}}$.\\
And,\\
\[{{\left[ \mathbf{C} \right]}_{:j}}=\frac{\partial f\left( \mathbf{z} \right)}{\partial {{x}_{j}}}=\sum\limits_{k\in \text{ind}\left( \mathbf{z} \right)}{\frac{\partial f\left( \mathbf{z} \right)}{\partial {{\left[ z \right]}_{k}}}\frac{\partial {{\left[ z \right]}_{k}}}{\partial {{x}_{j}}}}=\frac{\partial f\left( \mathbf{z} \right)}{\partial {{\mathbf{z}}^{\text{T}}}}\frac{\partial \mathbf{z}}{\partial {{x}_{j}}}\]
So, as a result\\
\[\frac{\partial \mathbf{y}}{\partial {{\mathbf{x}}^{\text{T}}}}=\frac{\partial f\left( \mathbf{z} \right)}{\partial {{\mathbf{z}}^{\text{T}}}}\frac{\partial \mathbf{z}}{\partial {{\mathbf{x}}^{\text{T}}}}\]

\subsection{Proof of  Ordinary Least Squares}
\[
  \underset{{{{{\bm \vartheta }}}}\in {{\mathbb{R}}^{N}}}{\mathop{\arg \min }}\, {{\left\| \bm \beta - \bm \Lambda {\bm \vartheta} \right\|}_{2}} \\
\]
\textbf{Proof:}\\
Suppose that $\mathbf y=\bm \beta - \bm \Lambda {\bm \vartheta}$, and the object function is $\mathbf y^{\text{T}} \mathbf y$.\\
And,\\
$\begin{aligned}
   \frac{\partial {{\mathbf{y}}^{\text{T}}}\mathbf{y}}{\partial \bm  \vartheta }&=\frac{\partial {{\mathbf{y}}^{\text{T}}}}{\partial \bm  \vartheta }\frac{\partial {{\mathbf{y}}^{\text{T}}}\mathbf{y}}{\partial \mathbf{y}}=\frac{\partial {{\mathbf{y}}^{\text{T}}}}{\partial \bm \vartheta }2\mathbf{y}=\frac{\partial {{\left(\bm  \beta -\mathbf{\Lambda }\bm \vartheta  \right)}^{\text{T}}}}{\partial \bm \vartheta }2\mathbf{y} \\
 & =\frac{\partial \left( {{\bm \beta }^{\text{T}}}-{{\bm \vartheta }^{\text{T}}}{{\mathbf{\Lambda }}^{\text{T}}} \right)}{\partial \bm \vartheta }2\mathbf{y}=-2{{\mathbf{\Lambda }}^{\text{T}}}\left(\bm  \beta -\mathbf{\Lambda }\bm \vartheta  \right)
\end{aligned}$ \\
So, as a result\\
$\frac{\partial {{\mathbf{y}}^{\text{T}}}\mathbf{y}}{\partial \bm  \vartheta }=\mathbf{0}\Rightarrow {{\mathbf{\Lambda }}^{\text{T}}}\bm \beta ={{\mathbf{\Lambda }}^{\text{T}}}\mathbf{\Lambda }\bm \vartheta \Rightarrow \bm \vartheta ={{\left( {{\mathbf{\Lambda }}^{\text{T}}}\mathbf{\Lambda } \right)}^{-\text{1}}}{{\mathbf{\Lambda }}^{\text{T}}}\bm \beta.$

\subsection{Proof of  Generalized Least Squares}
\[\underset{{{\bm \vartheta }}}{\mathop{\arg \min }}\,{{\left( {\bm {\beta }}-\mathbf{\Lambda }{{\bm \vartheta }} \right)}^{\text{T}}}{{\mathbf{\Omega }}^{-1}}\left( {{\bm \beta }}-\mathbf{\Lambda }{{\bm \vartheta }} \right)\]
\textbf{Proof:}\\
Suppose that $\mathbf y=\bm \beta - \bm \Lambda {\bm \vartheta}$, and the object function is $\mathbf y^{\text{T}} \mathbf{\Omega }^{-1} \mathbf y$.\\
And,

\[\begin{aligned}
  & \frac{\partial {{\mathbf{y}}^{\text{T}}}{{\mathbf{\Omega }}^{-1}}\mathbf{y}}{\partial  \bm \vartheta }=\frac{\partial {{\mathbf{y}}^{\text{T}}}}{\partial \bm  \vartheta }\frac{\partial {{\mathbf{y}}^{\text{T}}}{{\mathbf{\Omega }}^{-1}}\mathbf{y}}{\partial \mathbf{y}}={{\mathbf{\Lambda }}^{\text{T}}}\left( {{\mathbf{\Omega }}^{-1}}+{{\mathbf{\Omega }}^{-}}^{\text{T}} \right)\mathbf{y} \\
 & ={{\mathbf{\Lambda }}^{\text{T}}}\left( {{\mathbf{\Omega }}^{-1}}+{{\mathbf{\Omega }}^{-}}^{\text{T}} \right)\left( \bm  \beta -\mathbf{\Lambda } \bm \vartheta  \right)
\end{aligned}\]
So, as a result

\[\begin{aligned}
  & \frac{\partial {{\mathbf{y}}^{\text{T}}}{{\mathbf{\Omega }}^{-1}}\mathbf{y}}{\partial \bm \vartheta }=\mathbf{0}\Rightarrow {{\mathbf{\Lambda }}^{\text{T}}}\left( {{\mathbf{\Omega }}^{-1}}+{{\mathbf{\Omega }}^{-}}^{\text{T}} \right)\left(\bm  \beta -\mathbf{\Lambda }\bm \vartheta  \right)=\mathbf{0} \\
 & \Rightarrow {{\mathbf{\Lambda }}^{\text{T}}}\left( {{\mathbf{\Omega }}^{-1}}+{{\mathbf{\Omega }}^{-}}^{\text{T}} \right)\bm \beta ={{\mathbf{\Lambda }}^{\text{T}}}\left( {{\mathbf{\Omega }}^{-1}}+{{\mathbf{\Omega }}^{-}}^{\text{T}} \right)\mathbf{\Lambda }\bm \vartheta  \\
 & \Rightarrow \bm \vartheta ={{\left( {{\mathbf{\Lambda }}^{\text{T}}}\left( {{\mathbf{\Omega }}^{-1}}+{{\mathbf{\Omega }}^{-}}^{\text{T}} \right)\mathbf{\Lambda } \right)}^{-1}}{{\mathbf{\Lambda }}^{\text{T}}}\left( {{\mathbf{\Omega }}^{-1}}+{{\mathbf{\Omega }}^{-}}^{\text{T}} \right)\bm \beta  \\
\end{aligned}\]

\bibliographystyle{IEEEtran}
\bibliography{helx}

\begin{thebibliography}{10}
\providecommand{\url}[1]{#1}
\csname url@samestyle\endcsname
\providecommand{\newblock}{\relax}
\providecommand{\bibinfo}[2]{#2}
\providecommand{\BIBentrySTDinterwordspacing}{\spaceskip=0pt\relax}
\providecommand{\BIBentryALTinterwordstretchfactor}{4}
\providecommand{\BIBentryALTinterwordspacing}{\spaceskip=\fontdimen2\font plus
\BIBentryALTinterwordstretchfactor\fontdimen3\font minus
  \fontdimen4\font\relax}
\providecommand{\BIBforeignlanguage}[2]{{%
\expandafter\ifx\csname l@#1\endcsname\relax
\typeout{** WARNING: IEEEtran.bst: No hyphenation pattern has been}%
\typeout{** loaded for the language `#1'. Using the pattern for}%
\typeout{** the default language instead.}%
\else
\language=\csname l@#1\endcsname
\fi
#2}}
\providecommand{\BIBdecl}{\relax}
\BIBdecl

\bibitem{GAO1992Voltage}
B.~Gao, G.~K. Morison, and P.~Kundur, ``Voltage stability evaluation using
  modal analysis,'' \emph{Power Systems IEEE Transactions on}, vol.~7, no.~4,
  pp. 1529--1542, 1992.

\bibitem{Lugtu1980Power}
R.~L. Lugtu, D.~F. Hackett, K.~C. Liu, and D.~D. Might, ``Power system state
  estimation: Detection of topological errors,'' \emph{IEEE Power Engineering
  Review}, vol. PER-1, no.~1, pp. 19--19, 1980.

\bibitem{Wu1989Detection}
F.~F. Wu and W.~H.~E. Liu, ``Detection of topology errors by state estimation
  [power systems],'' \emph{IEEE Transactions on Power Systems}, vol.~4, no.~2,
  pp. 50--51, 1989.

\bibitem{he2015arch}
\BIBentryALTinterwordspacing
X.~He, Q.~Ai, R.~C. Qiu, W.~Huang, L.~Piao, and H.~Liu, ``A big data
  architecture design for smart grids based on random matrix theory,''
  \emph{IEEE Transactions on Smart Grid}, vol.~8, no.~2, pp. 674--686, 2017.
  [Online]. Available: \url{http://arxiv.org/pdf/1501.07329.pdf}
\BIBentrySTDinterwordspacing

\bibitem{Baek2015A}
J.~Baek, Q.~H. Vu, J.~K. Liu, X.~Huang, and X.~Yang, ``A secure cloud computing
  based framework for big data information management of smart grid,''
  \emph{IEEE Transactions on Cloud Computing}, vol.~3, no.~2, pp. 233--244,
  2015.

\bibitem{7364281}
Y.~C. Chen, J.~Wang, A.~D. Domínguez-García, and P.~W. Sauer,
  ``Measurement-based estimation of the power flow jacobian matrix,''
  \emph{IEEE Transactions on Smart Grid}, vol.~7, no.~5, pp. 2507--2515, Sept
  2016.

\bibitem{gomez2018electric}
A.~Gomez-Exposito, A.~J. Conejo, and C.~Canizares, \emph{Electric energy
  systems: analysis and operation}.\hskip 1em plus 0.5em minus 0.4em\relax CRC
  press, 2018.

\bibitem{Vaccaro2017A}
A.~Vaccaro and C.~A. Canizares, ``A knowledge-based framework for power flow
  and optimal power flow analyses,'' \emph{IEEE Transactions on Smart Grid},
  vol.~9, no.~1, pp. 230--239, 2017.

\bibitem{JD2012PowerDesign}
G.~JD and S.~MS, \emph{Power System Analysis and Design}.\hskip 1em plus 0.5em
  minus 0.4em\relax Boston, MA, USA: Cengage, 2012.

\bibitem{van1991total}
S.~Van~Huffel and J.~Vandewalle, \emph{The total least squares problem:
  computational aspects and analysis}.\hskip 1em plus 0.5em minus 0.4em\relax
  Siam, 1991, vol.~9.

\bibitem{wiki2018TLS}
\BIBentryALTinterwordspacing
{Wikipedia}, ``Total least squares,'' 2018. [Online]. Available:
  \url{https://en.wikipedia.org/wiki/Total_least_squares}
\BIBentrySTDinterwordspacing

\bibitem{strutz2010data}
T.~Strutz, \emph{Data fitting and uncertainty: A practical introduction to
  weighted least squares and beyond}.\hskip 1em plus 0.5em minus 0.4em\relax
  Vieweg and Teubner, 2010.

\bibitem{wiki2018kron}
\BIBentryALTinterwordspacing
{Wikipedia}, ``Kronecker product,'' 2018. [Online]. Available:
  \url{https://en.wikipedia.org/wiki/Kronecker_product}
\BIBentrySTDinterwordspacing

\bibitem{MATPOWER2011matpower}
R.~Zimmerman, C.~Murillo-S{\'a}nchez, and D.~Gan, ``\text{Matpower User{\'s}
  Manual, Version 4.1},'' \emph{Power Systems Engineering Research Center},
  2011.

\bibitem{he2017invisible}
\BIBentryALTinterwordspacing
H.~Xing, R.~C. Qiu, C.~Lei, A.~Qian, Z.~Ling, and Z.~Jian, ``Detection and
  estimation of the invisible units using utility data based on random matrix
  theory,'' \emph{ArXiv e-prints}, 2017. [Online]. Available:
  \url{http://arxiv.org/pdf/1710.10745.pdf}
\BIBentrySTDinterwordspacing

\end{thebibliography}

\normalsize{}
\end{document}